\documentclass[
 reprint,
 amsmath,amssymb,
 aps, onecolumn, nofootinbib, 12pt, a4paper
]{revtex4-2}

\usepackage{xcolor}
\usepackage{graphicx}
\usepackage{dcolumn}
\usepackage{bm}
\usepackage{hyperref}
\hypersetup{colorlinks, allcolors=blue}
\usepackage{physics}
\usepackage{bm}
\usepackage{booktabs}
\usepackage[caption=false]{subfig}

\newcolumntype{C}[1]{>{\centering\arraybackslash}p{#1}}

\newcommand{\Lagr}{\mathcal{L}}
\newcommand{\bigO}{\mathcal{O}}
\newcommand{\VM}{\mathrm{VM}}
\newcommand{\DM}{\mathrm{DM}}
\newcommand{\qL}{\mu_{q_L}}
\newcommand{\lL}{\mu_{l_{iL}}}
\newcommand{\eR}{\mu_{e_{iR}}}
\newcommand{\uR}{\mu_{u_{iR}}}
\newcommand{\dR}{\mu_{d_{iR}}}

\newcommand{\mphiv}{\mu_{\Phi}}

\newcommand{\uRd}{\mu_{u_R'}}
\newcommand{\dRd}{\mu_{d_R'}}
\newcommand{\nRd}{\mu_{\nu_{iR}'}}
\newcommand{\eRd}{\mu_{e_{iR}'}}
\newcommand{\eL}{\mu_{e_{iL}'}}
\newcommand{\uL}{\mu_{u_{iL}'}}
\newcommand{\dL}{\mu_{d_{iL}'}}

\newcommand{\mpOd}{\mu_{{\phi^0}'}}
\newcommand{\mppd}{\mu_{{\phi^+}'}}
\newcommand{\VASB}{V_{\mathrm{ASB}}}
\newcommand{\Vfull}{V_{\mathrm{M2HDM}}}

\begin{document}

\title{Implementing asymmetric dark matter and dark electroweak baryogenesis in a mirror two-Higgs-doublet model}

\author{Alexander C. Ritter}
 \email{rittera@student.unimelb.edu.au (corresponding author)}
\author{Raymond R. Volkas}
 \email{raymondv@unimelb.edu.au}
\affiliation{
ARC Centre of Excellence for Dark Matter Particle Physics,
School of Physics, \\
The University of Melbourne,
Victoria 3010, Australia
}

\begin{abstract}
Models of asymmetric dark matter (ADM) seek to explain the apparent coincidence between the present-day mass densities of visible and dark matter, $\Omega_{\DM} \simeq 5\Omega_{\VM}$. However, most ADM models only relate the number densities of visible and dark matter without motivating the similar particle masses. We expand upon a recent work that obtained a natural mass relationship in a mirror matter ADM model with two Higgs doublets in each sector, by looking to implement dark electroweak baryogenesis as the means of asymmetry generation. We explore two aspects of the mechanism: the nature of the dark electroweak phase transition, and the transfer of particle asymmetries between the sectors by the use of portal interactions. We find that both aspects can be implemented successfully for various regions of the parameter space. We also analyse one portal interaction -- the neutron portal -- in greater detail, in order to satisfy the observational constraints on dark radiation.
\end{abstract}

\maketitle

\section{Introduction}
Determining the particle nature of dark matter (DM) remains one of the most important problems in fundamental physics. While there are some important constraints on its nature -- for example, it cannot be hot DM because large-scale structure formation then yields incorrect results -- DM is famous, or notorious, for being anything from ``fuzzy'' scalars at the $10^{-22}$ eV mass scale \cite{Hu:2000ke}, to several solar-mass primordial black holes \cite{Carr:2020xqk}, with many different kinds of possibilities at intermediate mass scales \cite{Feng:2010gw}. It therefore makes sense to carefully examine what we \emph{do} know observationally about DM, because there may be clues already lurking in the data about what its fundamental nature is.

One fact that may be important is the apparent coincidence in the present-day cosmological mass densities of visible and dark matter, which obey
\begin{equation}
    \Omega_{\textrm{DM}} \simeq 5\, \Omega_{\textrm{VM}},
    \label{eq:Omega-coincidence}
\end{equation}
where $\Omega_X$ is the mass density of $X$ divided by the critical density \cite{Aghanim:2018eyx}. Cosmologically one would expect different relic species to have very different mass and/or number densities unless there are fundamental reasons for it to be otherwise. For example, the equal number densities of protons and electrons is a consequence of the basic requirement of electric charge neutrality for the universe. It is thus worth exploring the hypothesis that Eq.~(\ref{eq:Omega-coincidence}) is the result of a deep connection between visible and dark matter rather than being a true coincidence.

For most DM candidates, the physics determining the relic density -- for example, the freeze out process for a thermal relic -- has no connection with the physics driving the mass density of visible matter: baryogenesis, which sets the proton number density, and the confinement scale of quantum chromodynamics (QCD), which sets the proton mass. 

Asymmetric DM (ADM) is an exception to this general rule, since the relic number density of DM particles is then determined by an asymmetry in the dark sector that is chemically related to the baryon asymmetry. Asymmetric DM is a paradigm, and many different models have been proposed (for reviews, see \cite{Davoudiasl:2012uw,Petraki:2013wwa,Zurek:2013wia}). The vast majority of these proposals provide specific dynamics to relate the number density asymmetries, but are silent on why the DM mass seems apparently to be related to the proton mass. Yet without such a connection, ADM models fail to explain the cosmological coincidence. Instead the factor of five in Eq.~(\ref{eq:Omega-coincidence}) is used to ``predict'' the DM mass within schemes that only relate the number densities. Clearly, this is unsatisfactory. The purpose of this paper is to continue analysing ways to connect the DM and proton masses within an ADM model. Note that it is not our goal to explain exactly why the approximate ratio is the specific value of five. Rather, our goal is to construct a theory where a ratio of order one is relatively generic, and the precise value can be fitted by choosing parameters appropriately.

We are faced with the task of explaining why the DM mass should have anything to do with the confinement scale of QCD. The most obvious idea is that DM is a baryonlike bound state of an interaction in the dark sector that resembles QCD. Several such schemes have been proposed and analysed in the literature, though only a few of them have the DM-proton mass connection as a motivation \cite{Nussinov:1985xr, Barr:1990ca, Hodges:1993yb, Foot:2004pa, Chacko:2005pe, Kribs:2009fy, An:2009vq, Frandsen:2010yj, Cline:2013zca, Appelquist:2013ms, Farina:2015uea, Garcia:2015toa, Farina:2016ndq, Beauchesne:2020mih, Bai:2013xga, Newstead:2014jva, Hall:2019rld}. Indeed, with an arbitrary confining gauge force in the dark sector and a general particle content, there is no reason for the confinement scale to be near that of visible QCD. Two exceptions to this have been proposed: (i) the dark gauge group mirrors QCD by being SU(3), and the two confinement scales are related through either an exact or somewhat broken symmetry that connects the two SU(3) sectors, and (ii) the particle content is chosen so that the two running coupling constants approach infrared fixed points whose magnitudes are similar \cite{Bai:2013xga, Newstead:2014jva}. Both ideas have merit, and in this paper we consider option (i).

\emph{A priori}, the two SU(3) sectors could be related by either a continuous or a discrete symmetry. But it is difficult to make the former work, because of the necessary appearance of multiplets that transform under both the dark SU(3) and the usual standard model (SM) gauge group $G_{\textrm{SM}}$. We therefore focus on the discrete symmetry possibility, specifically the simplest case of $Z_2$. We seek a theory where the DM is a baryonic bound state of ``dark quarks'' which are triplets under dark SU(3) and singlets under $G_{\textrm{SM}}$. Because the usual quarks are in the $(\mathbf{3},\mathbf{2},\frac{1}{6})$, $(\mathbf{3},\mathbf{1},\frac{2}{3})$ and $(\mathbf{3},\mathbf{1},-\frac{1}{3})$ representations of $G_{\textrm{SM}}$, the only way their $Z_2$ partners can be singlets under SM forces is if we duplicate the electroweak sector as well. We are evidently driven to a mirror-matter type of gauge group,
\begin{equation}
    G \times G'
\end{equation}
where $G'$ is isomorphic to $G$ with prime denoting the dark sector.

In this paper we continue in the vein of Ref.~\cite{Lonsdale:2018xwd} and consider the simplest case where $G$ is just the SM gauge group SU(3)$\times$SU(2)$\times$U(1), which is exactly the mirror-symmetric extension of the SM. The $Z_2$ symmetry interchanges the visible and dark sectors and enforces equal QCD and dark-QCD coupling constants when it is exact. To relate the DM mass to the proton mass we need some kind of a connection between the two QCD coupling constants. This connection may be the strict equality of the coupling constants and hence also the confinement scales, or some $Z_2$ breaking can be introduced so as to remove the exact equality but retain a relationship. The unbroken case has been extensively studied -- see, for example, Refs. \cite{Lee:1956qn, Kobzarev:1966qya, Pavsic:1974rq, Blinnikov:1982eh, Foot:1991bp, Foot:1991py, Hodges:1993yb, Foot:1995pa, Berezhiani:2000gw, Bento:2002sj, Berezhiani:2003xm, Ignatiev:2003js, Foot:2003iv, Foot:2003jt, Foot:2004pq, Foot:2004pa, Ciarcelluti:2004ik, Ciarcelluti:2004ip, Chacko:2005pe, Foot:2010hu, Foot:2014mia, Cerulli:2017jzz}. We choose to follow Ref.~\cite{Lonsdale:2018xwd} and explore a spontaneously broken $Z_2$ scenario (see also Refs. \cite{Berezhiani:1996sz, Foot:2000tp, Berezhiani:2008gi,An:2009vq, Cui:2011wk, Gu:2012fg, Addazi:2015cua, Farina:2015uea, Garcia:2015toa, Farina:2016ndq, Beauchesne:2020mih}). Part of the motivation for that is to permit the DM candidate to be a single dark-neutronlike particle, rather than having to deal with the complicated (though very interesting) situation of exact mirror-DM. 

Reference \cite{Lonsdale:2018xwd} was based on the process of ``asymmetric symmetry breaking (ASB).'' This is a spontaneous symmetry breaking scheme that permits the two sectors to break quite differently despite the $Z_2$ symmetry of the Lagrangian. It is distinct from the idea of introducing a $Z_2$-odd scalar whose vacuum expectation value (VEV) breaks the mirror symmetry, in that in general it affords more flexibility in the symmetry-breaking outcome.\footnote{The ASB mechanism really comes into its own when you want to break $G$ and $G'$ to \emph{different subgroups}. This will not be the case in the model analysed in this paper. However, the mirror-symmetric SM, rather than a mirror-symmetric theory with an extended gauge group such as a grand unified theory, permits us to focus on the DM physics rather than being distracted by the many unrelated issues that arise from SM gauge-group extensions. Ultimately, the proper context for ASB may well be a grand unified theory, as discussed in the original papers \cite{Lonsdale:2014wwa,Lonsdale:2014yua}. The mirror-symmetric SM analysed here could then be the low-energy effective theory of a more ambitious model.} Reference \cite{Lonsdale:2018xwd} analysed a mirror-symmetric model with two Higgs doublets in each sector. The ASB process was then employed to ensure that the doublets, one from each sector, that gain the dominant VEVs are \emph{not} $Z_2$ partners. This allows the dark-fermion masses and mass ratios to be completely different from the usual quark and lepton masses and ratios. Reference \cite{Lonsdale:2018xwd} described an attempt at a full theory that saw visible and dark baryogenesis occur through the familiar sphaleron-reprocessed type I seesaw leptogenesis dynamics driven by the out-of-equilibrium decays of heavy neutral leptons. 

The purpose of the present paper is:
\begin{itemize}
    \item To construct an alternative version of the theory where asymmetry generation occurs through dark electroweak baryogenesis, which is another reasonable mechanism that is worth exploring. We show that there is sufficient freedom to arrange for the dark electroweak phase transition to be strongly first order, as required for this mechanism. Such a phenomenon may give rise to gravitational waves that are detectable through future space-based interferometers \cite{Caprini:2019egz}.
    \item To analyse some minimal possibilities for how the dark asymmetry may be reprocessed into a visible asymmetry through various higher dimension portal interactions\footnote{For previous applications of asymmetry transfer through effective operator portal interactions in ADM models, see \cite{Foot:2004pq,Shelton:2010ta,Feng:2013wn}. We also note Ref.~\cite{Hall:2019rld}, in which the renormalisable neutrino portal is responsible for asymmetry transfer following dark electroweak baryogenesis in a dark sector with a mirror gauge group and two Higgs doublets.}. We also discuss the generation of these asymmetries, and contend that there should be sufficient $CP$ violation in the dark CKM matrix to produce the required baryon asymmetry.
    \item To continue to analyse the quite difficult problem of how observational constraints on dark radiation may be obeyed in such a theory. One of our goals here is to present a clear account of the challenges in achieving this aim without introducing fine-tuning that is as bad or worse than the cosmological coincidence puzzle of Eq.~\ref{eq:Omega-coincidence}.
\end{itemize}

The remainder of this paper is structured as follows: In Sec.~\ref{sec:model} we outline the model and provide some theoretical and experimental constraints on the parameters of the theory. These will help guide our search in Sec.~\ref{sec:EWPT}, where we analyse the dynamics of the dark electroweak phase transition and identify areas of parameter space for which the transition is strongly first-order. Such a transition is necessary to allow for the generation of a dark baryon asymmetry through electroweak baryogenesis. In Sec.~\ref{sec:asym} we discuss the generation of this asymmetry and consider its partial reprocessing into a visible baryon asymmetry through a number of effective operator portal interactions. In Sec.~\ref{sec:darkrad} we then analyse one of these possibilities -- the ``neutron portal'' -- in more detail, as it can also play a role in avoiding strong observational bounds on additional dark radiation. This introduces a number of difficulties, which we clearly outline, before providing some concluding remarks in Sec.~\ref{sec:conclusion}.

\section{The Model and Constraints}\label{sec:model}
As this work builds off the mirror two Higgs doublet model of Ref.~\cite{Lonsdale:2018xwd}, we do not provide a fully detailed description of the theory in this section. Rather, we summarise the salient details of the model so that the contents of this paper can be understood in isolation, as well as highlighting the points where we differ. We also provide some more specific restraints on the parameters of the model, and in particular on the couplings and mass terms in the scalar potential. These are especially relevant for Sec.~\ref{sec:EWPT}, where we determine whether the model can accommodate a strong first-order dark electroweak phase transition; the restrictions we discuss here will help guide our search through the large parameter space of the scalar sector.

The gauge group is SU(3)$\times$SU(2)$\times$U(1)$\times$SU(3)$'\times$SU(2)$'\times$U(1)$'$, where the mirror (dark) sector is a duplicated version of the standard model. The dark gauge groups and particles are indicated by primes. The dark particle content is a copy of the visible particle content, as required by a discrete $Z_2$ parity symmetry that exchanges SM particles with their dark counterparts. The particle transformation properties are given by
\begin{equation}\label{eq:mirror-transforms}
    \phi \leftrightarrow \phi', \quad G^\mu \leftrightarrow G_\mu', \quad f_L \leftrightarrow f_R', \quad f_R \leftrightarrow f_L',
\end{equation}
where $\phi$, $G^\mu$, and $f$ are scalar, gauge, and fermion fields respectively. Note that as a parity symmetry, the $Z_2$ exchanges left-handed and right-handed particles. (The chirality flip feature is an aesthetic choice, and is not essential.)

\begin{table}
\begin{tabular}{C{1cm}C{3cm}|C{1cm}C{3cm}}
\toprule\hline
$q_{iL}$ & $(\mathbf{3}, \mathbf{2}, -\frac{1}{6})(\mathbf{1}, \mathbf{1}, 0)$ & $q_{iR}'$ & $(\mathbf{1}, \mathbf{1}, 0)(\mathbf{3}, \mathbf{2}, -\frac{1}{6})$ \\
$u_{iR}$ & $(\mathbf{3}, \mathbf{1}, \frac{2}{3})(\mathbf{1}, \mathbf{1}, 0)$ & $u_{iL}'$ & $(\mathbf{1}, \mathbf{1}, 0)(\mathbf{3}, \mathbf{1}, \frac{2}{3})$ \\
$d_{iR}$ & $(\mathbf{3}, \mathbf{1}, -\frac{1}{3})(\mathbf{1}, \mathbf{1}, 0)$ & $d_{iL}'$ & $(\mathbf{1}, \mathbf{1}, 0)(\mathbf{3}, \mathbf{1}, -\frac{1}{3})$ \\
$l_{iL}$ & $(\mathbf{1}, \mathbf{2}, -\frac{1}{2})(\mathbf{1}, \mathbf{1}, 0)$ & $l_{iR}'$ & $(\mathbf{1}, \mathbf{1}, 0)(\mathbf{1}, \mathbf{2}, -\frac{1}{2})$ \\
$e_{iR}$ & $(\mathbf{1}, \mathbf{1}, -1)(\mathbf{1}, \mathbf{1}, 0)$ & $e_{iL}'$ & $(\mathbf{1}, \mathbf{1}, 0)(\mathbf{1}, \mathbf{1}, -1)$ \\
\hline
$\Phi_1$ & $(\mathbf{1}, \mathbf{2}, 0)(\mathbf{1}, \mathbf{1}, 0)$ & $\Phi_1'$ & $(\mathbf{1}, \mathbf{1}, 0)(\mathbf{1}, \mathbf{2}, 0)$ \\
$\Phi_2$ & $(\mathbf{1}, \mathbf{2}, 0)(\mathbf{1}, \mathbf{1}, 0)$ & $\Phi_2'$ & $(\mathbf{1}, \mathbf{1}, 0)(\mathbf{1}, \mathbf{2}, 0)$ \\
\hline\bottomrule
\end{tabular}
\caption{\label{table:particlecontent}The particle content and their representations under the mirror symmetric gauge group (SU(3)$\times$SU(2)$\times$U(1))$\times$(SU(3)$'\times$SU(2)$'\times$U(1)$'$)}
\end{table}

The total particle content of the model is given in Table \ref{table:particlecontent}. The fermion content consists of the standard model fermions and their dark partners. Note that unlike the original paper, we do not list right-handed singlet neutrinos and their partners. These were introduced to allow for asymmetry generation through thermal leptogenesis, whereas we will be considering dark electroweak baryogensis as the asymmetry creation mechanism.\footnote{Of course, we need to generate massive neutrinos somehow, but we may remain largely agnostic about the precise mechanism for present purposes, only requiring that it not dominate asymmetry generation and also not contribute significantly to washout.} There are four scalars in the model: two Higgs doublets, $\Phi_1$ and $\Phi_2$, along with their dark counterparts $\Phi_1'$ and $\Phi_2'$. The additional Higgs doublets allow for the ASB mechanism \cite{Lonsdale:2014wwa} to be implemented; as a vital component of the model, understanding the mechanism will be central to constructing the scalar potential.

\subsection{The scalar potential and asymmetric symmetry breaking}

To introduce the ASB mechanism we first consider the scalar potential in an illustrative toy model. In addition to the mirror $Z_2$ symmetry exchanging $\Phi_1$ and $\Phi_2$ with $\Phi_1'$ and $\Phi_2'$, we impose extra discrete $Z_2$ symmetries such that only terms with even numbers of a given scalar are allowed. Then, the scalar potential can be written in the form
\begin{equation} \label{Eq:VASB}
\begin{split}
    \VASB &= \lambda_1\left(\Phi_1^\dagger\Phi_1 + \Phi_1'^\dagger\Phi_1'-\frac{v^2}{2}\right)^2 +\lambda_2\left(\Phi_2^\dagger\Phi_2 + \Phi_2'^\dagger\Phi_2'-\frac{w^2}{2}\right)^2 \\
    &+ \kappa_1\left(\Phi_1^\dagger\Phi_1\right)\left(\Phi_1'^\dagger\Phi_1'\right)
    + \kappa_2\left(\Phi_2^\dagger\Phi_2\right)\left(\Phi_2'^\dagger\Phi_2'\right) \\
    &+ \sigma_1\left(\left(\Phi_1^\dagger\Phi_1\right)\left(\Phi_2^\dagger\Phi_2\right) + \left(\Phi_1'^\dagger\Phi_1'\right)\left(\Phi_2'^\dagger\Phi_2'\right)\right) \\
    &+ \sigma_2\left(\Phi_1^\dagger\Phi_1 + \Phi_1'^\dagger\Phi_1' + \Phi_2^\dagger\Phi_2 + \Phi_2'^\dagger\Phi_2'-\frac{v^2}{2}-\frac{w^2}{2}\right)^2 .
\end{split}
\end{equation}

In the parameter space region where each of $[\lambda_1, \lambda_2, \kappa_1, \kappa_2, \sigma_1, \sigma_2]$ are positive, the global minimum occurs when all terms are independently zero. This can be achieved by the following pattern of VEVs:
\begin{equation}
\begin{split}
    \ev{\Phi_1} = 
    \begin{bmatrix}
    0\\
    \frac{v}{\sqrt{2}}
    \end{bmatrix},
    \quad
    \ev{\Phi_1'} = 0,
    \\
    \ev{\Phi_2} = 0,
    \quad
    \ev{\Phi_2'} = 
    \begin{bmatrix}
    0\\
    \frac{w}{\sqrt{2}}
    \end{bmatrix}.
\end{split}
\end{equation}
This minimum clearly breaks the mirror $Z_2$ symmetry, with non-mirror partner Higgs doublets gaining nonzero VEVs in the two sectors. The motivation for breaking the mirror symmetry is to obtain differing particle masses in the visible and dark sectors. As the masses of the visible and dark baryons result from the QCD confinement energy of the SU(3) interaction in each sector, we want the QCD confinement scale in each sector -- $\Lambda_{\mathrm{QCD}}$ and $\Lambda_{\DM}$ -- to differ. 

This was explored in Ref. \cite{Lonsdale:2014wwa}, which considered the evolution of the SU(3) gauge couplings $\alpha_3$ and $\alpha_3'$. At temperatures above the scale of mirror symmetry breaking, these couplings are equal; after the mirror symmetry is broken by the development of an asymmetric minimum, their running to low energies depends upon the spectrum of quark masses in each sector. Thanks to the asymmetric symmetry breaking minimum, these two spectra are independent, as the Higgs doublets that give masses to the quarks in each sector are not mirror partners. If the minimum is constructed such that $w \gg v$, then depending on the Yukawa couplings of the quarks to $\Phi_2'$, a dark confinement scale $\Lambda_{\DM}$ a factor of a few higher than $\Lambda_{\mathrm{QCD}}$ can be easily achieved. This is encapsulated in Fig. 1 of Ref.~\cite{Lonsdale:2018xwd}, which plots $\Lambda_{\DM}$ against the ratio between electroweak scales ($\rho \equiv w/v$) for a selection of dark quark mass spectra. 

While an asymmetric symmetry breaking minimum can be readily obtained for the toy scalar potential, the situation is more complex when the full scalar potential is considered. The most general mirror two-Higgs-doublet scalar potential -- where we only impose the mirror $Z_2$ exchange symmetry -- is given by
\begin{equation} \label{Eq:Vfull}
\begin{split}
    \Vfull &= m_{11}^2\left(\Phi_1^\dagger\Phi_1 + \Phi_1'^\dagger\Phi_1'\right) + m_{22}^2\left(\Phi_2^\dagger\Phi_2 + \Phi_2'^\dagger\Phi_2'\right) \\
    &+ \left(m_{12}^2\left(\Phi_1^\dagger\Phi_2 + \Phi_1'^\dagger\Phi_2'\right) + H.c.\right) 
    + \frac{1}{2}z_1\left(\left(\Phi_1^\dagger\Phi_1\right)^2 + \left(\Phi_1'^\dagger\Phi_1'\right)^2\right) \\
    &+ \frac{1}{2}z_2\left(\left(\Phi_2^\dagger\Phi_2\right)^2 + \left(\Phi_2'^\dagger\Phi_2'\right)^2\right) + z_3\left(\Phi_1^\dagger\Phi_1\Phi_2^\dagger\Phi_2 + \Phi_1'^\dagger\Phi_1'\Phi_2'^\dagger\Phi_2'\right) \\
    &+ z_4\left(\Phi_1^\dagger\Phi_2\Phi_2^\dagger\Phi_1 + \Phi_1'^\dagger\Phi_2'\Phi_2'^\dagger\Phi_1'\right) + \frac{1}{2}z_5\left(\left(\Phi_1^\dagger\Phi_2\right)^2 + \left(\Phi_1'^\dagger\Phi_2'\right)^2 + H.c.\right) \\
    &+ \left[\left(z_6\Phi_1^\dagger\Phi_1 + z_7\Phi_2^\dagger\Phi_2\right)\Phi_1^\dagger\Phi_2 + \left(z_6\Phi_1'^\dagger\Phi_1' + z_7\Phi_2'^\dagger\Phi_2'\right)\Phi_1'^\dagger\Phi_2' + H.c.\right] \\
    &+ z_8\Phi_1^\dagger\Phi_1\Phi_1'^\dagger\Phi_1' + z_9\Phi_2^\dagger\Phi_2\Phi_2'^\dagger\Phi_2' + \left(z_{10}\Phi_1^\dagger\Phi_2\Phi_1'^\dagger\Phi_2' + H.c.\right) \\
    &+ \left(z_{11}\Phi_1^\dagger\Phi_2\Phi_2'^\dagger\Phi_1' + H.c.\right) + z_{12}\left(\Phi_1^\dagger\Phi_1\Phi_2'^\dagger\Phi_2' + \Phi_1'^\dagger\Phi_1'\Phi_2^\dagger\Phi_2\right) \\
    &+ \left[\left(z_{13}\Phi_1^\dagger\Phi_1 + z_{14}\Phi_2^\dagger\Phi_2\right)\Phi_1'^\dagger\Phi_2' + \left(z_{13}\Phi_1'^\dagger\Phi_1' + z_{14}\Phi_2'^\dagger\Phi_2'\right)\Phi_1^\dagger\Phi_2 + H.c.\right].
\end{split}
\end{equation}

The large number of new terms prevents us from constructing the potential in such a way that its minimum exactly follows the asymmetric symmetry breaking pattern of Eq.~\ref{Eq:VASB}. In general, the global minimum is given by
\begin{equation}
\ev{\Phi_i} = 
\begin{bmatrix}
0\\
\frac{v_i}{\sqrt{2}}
\end{bmatrix},
\quad
\ev{\Phi_i'} = 
\begin{bmatrix}
0\\
\frac{w_i}{\sqrt{2}}
\end{bmatrix}
\end{equation}
meaning that all four doublets have nonzero VEVs. To recover the pattern of Eq.~\ref{Eq:VASB}, we transform to a basis in which one Higgs doublet in each sector has a zero VEV. This ``dual Higgs basis'' is defined by 
\begin{equation}\label{Eq:newbasis}
\begin{split}
    H_1 = \frac{v_1^*\Phi_1 + v_2^*\Phi_2}{v}, 
    \quad 
    H_2 = \frac{-v_2\Phi_1 + v_1\Phi_2}{v},
    \\
    H_1' = \frac{w_1^*\Phi_1' + w_2^*\Phi_2'}{w},
    \quad
    H_2' = \frac{-w_2\Phi_1' + w_1\Phi_2'}{w},
\end{split}
\end{equation}
where
\begin{equation}
    v = \sqrt{\abs{v_1}^2 + \abs{v_2}^2}, \quad w = \sqrt{\abs{w_1}^2 + \abs{w_2}^2}.
\end{equation}
With these assignments, only $H_1$ and $H_1'$ gain non-zero VEVs, and they will not be mirror partners if we have $v_1 \neq w_1$ and $v_2 \neq w_2$. 

We wish to maintain the desirable features of exact asymmetric symmetry breaking, where unrelated Higgs bosons are responsible for mass generation in each sector. Thus, we want $H_1$ and $H_1'$ to be largely independent admixtures of $\Phi_1^{(\prime)}$ and $\Phi_2^{(\prime)}$. This can be achieved by a global minimum that resembles the ASB minimum; that is, one where
\begin{equation}\label{Eq:asb_limit}
    v_1 \gg v_2, \quad w_1 \ll w_2, \quad w_2 \gg v_1.
\end{equation}
We will refer to this as the ``ASB limit.'' 

To obtain it, we want to choose the parameters for $\Vfull$ such that the potential is of a similar form to $\VASB$. Equating coefficients between the two potentials, we obtain 
\begin{equation}\label{Eq:param_relationships}
\begin{split}
    {m_{11}}^2 = -\lambda_1v^2 -\sigma_2(v^2+w^2), \quad z_1 = 2\lambda_1 + 2\sigma_2, &\quad z_8 = \kappa_1 + 2\lambda_1 + 2\sigma_2,\\
    {m_{22}}^2 = -\lambda_2w^2 -\sigma_2(v^2+w^2), \quad z_2 = 2\lambda_2 + 2\sigma_2, &\quad z_9 = \kappa_2 + 2\lambda_2 + 2\sigma_2,\\
    z_3 = \sigma_1 + 2\sigma_2, &\quad z_{12} = 2\sigma_2.\\
\end{split}
\end{equation}
The other parameters in $\Vfull$ do not correspond with any terms in $\VASB$. Thus, to approximately replicate the form of $\VASB$ in the full potential, we can initially apply a rough condition that these additional parameters are small with respect to those listed above; that is,
\begin{equation}
    z_1, z_2, z_3, z_8, z_9, z_{12} \gg z_4, z_5, z_6, z_7, z_{10}, z_{11}, z_{13}, z_{14}.
\end{equation}

\begin{table}[t]
\centering
\begin{tabular}{C{5cm}C{5cm}} 
\toprule
\textbf{Parameters} & \textbf{Values}     \\ 
\hline
${m_{11}}^2$ & $-(87~\mathrm{GeV})^2$    \\
${m_{12}}^2$ & $-(90~\mathrm{GeV})^2$     \\
${m_{22}}^2$ & $-(2600~\mathrm{GeV})^2$    \\
$z_1$, $z_2$      & 0.129      \\
$z_3$, $z_8$, $z_9$, $z_{10}$      & 0.8        \\
$z_4$, $z_5$, $z_6$, $z_7$, $z_{11}$, $z_{13}$, $z_{14}$      & 0.01       \\
$z_{12}$      & $1\times10^{-8}$  \\
\bottomrule
\end{tabular}
\caption{Benchmark parameter point for the full scalar potential, taken from Table 1 in Ref.~\cite{Lonsdale:2018xwd}.}\label{table:param1}
\end{table}

To see how this might be applied, we consider the benchmark parameter point given in Table 1 of the original paper \cite{Lonsdale:2018xwd}, which we reproduce in Table \ref{table:param1}. Most of the parameters satisfy our rough condition, with two notable exceptions. First, $z_{12}$ is by far the smallest coupling; this will be motivated when we consider the scalar masses of the theory. In addition, $z_{10}$ is just as large as the other quartic couplings. Even though it does not correspond to any terms in the toy model potential, making $z_{10}$ large does not alter the asymmetric symmetry breaking pattern of the minimum; the term $(z_{10}\Phi_1^\dagger\Phi_2\Phi_1'^\dagger\Phi_2' + H.c.)$ contains both Higgs bosons that gain small VEVs, and thus only provides a small contribution to the potential at the asymmetric symmetry breaking minimum. By this logic, $z_4$, $z_5$, and $z_{11}$ also do not necessarily have to be small. Thus, ensuring an asymmetric symmetry breaking pattern for the minimum of $\Vfull$ only requires
\begin{equation}\label{Eq:smallzs}
    z_1, z_2, z_3, z_8, z_9 \gg z_6, z_7, z_{11}, z_{13}, z_{14}.
\end{equation}
This is rather a rough condition, and there will be more nuance in exactly how small these quartic couplings will need to be. 

To conclude this discussion we note what happens for large values of the dark electroweak scale $w$. As can be seen in Table \ref{table:param1}, a valid parameter point can be achieved with only one to two orders of magnitude difference between the couplings (except for the aforementioned $z_{12}$). This benchmark point corresponds to $w = 7276$ GeV, thirty times greater than the visible VEV $v = 246$ GeV. For values of $w$ one or more orders of magnitude larger than this, an issue will arise from the term $(z_{14}\Phi_1^\dagger\Phi_2\Phi_2'^\dagger\Phi_2' + h.c.)$. Expanding around the VEV of $\Phi_2'$ we obtain the term $(z_{14}{w_2}^2)\Phi_1^\dagger\Phi_2$, which will strongly alter the tree-level ASB minimum when $z_{14}{w_2}^2$ is of the order of the larger scalar couplings. To preserve the asymmetric symmetry breaking pattern as we increase the value of $w$, $z_{14}$ must then be made smaller than $0.01$. 

\subsection{Scalar masses}

In deriving the masses of the scalar content of the model, we follow the original paper and define the field content of the doublets by
\begin{equation} \label{Eq:phifields}
\begin{split}
    \Phi_1 = 
    \begin{bmatrix}
    G_1^+\\
    \frac{1}{\sqrt{2}}(v_1 + \phi_1 + iG_1)
    \end{bmatrix},&
    \quad
    \Phi_1' = 
    \begin{bmatrix}
    I_1^+ \\
    \frac{1}{\sqrt{2}}(w_1 + \phi_1' + ia_1)
    \end{bmatrix},
    \\
    \Phi_2 = 
    \begin{bmatrix}
    I_2^+ \\
    \frac{1}{\sqrt{2}}(v_2 + \phi_2 + ia_2)
    \end{bmatrix},&
    \quad
    \Phi_2' = 
    \begin{bmatrix}
    {G_2^+}'\\
    \frac{1}{\sqrt{2}}(w_2 + {\phi_2}' + iG_2)
    \end{bmatrix}.
\end{split}
\end{equation}
Generically we obtain a 16 $\times$ 16 mass matrix which produces 10 nonzero mass eigenstates when diagonalised. When working with real parameters in $\Vfull$, there will only be mixing between fields at the same position in each doublet; thus, we only have to diagonalise four separate 4 $\times$ 4 mass matrices. These four matrices each involve mixing between visible and dark fields, and thus the mass eigenstates are generically admixtures of visible and dark states. 

This is an issue, as we require one of these states to serve as the SM Higgs boson $h$, which must not mix strongly with any of the new scalars. We must especially avoid any dependence of the mass of $h$ on the dark scale $w$. To see how this is achieved, we consider the mass mixing matrix for the $\phi$ fields. These terms derive from Appendix A in the original paper, where we work in the ASB limit for the minimum; that is, we ignore terms involving the small VEVs $v_2$ and $w_1$, and work only with real quartic couplings. We then obtain the mass matrix
\begin{equation}
    \frac{1}{2}\left(\phi_1, \phi_2, \phi_1', \phi_2'\right)
    \begin{pmatrix}
    m_{\phi_1\phi_1} & m_{\phi_1\phi_2} & m_{\phi_1\phi_1'} & m_{\phi_1\phi_2'} \\
    m_{\phi_1\phi_2} & m_{\phi_2\phi_2} & m_{\phi_1\phi_2'} & m_{\phi_2\phi_2'} \\
    m_{\phi_1\phi_1'} & m_{\phi_1\phi_2'} & m_{\phi_1'\phi_1'} & m_{\phi_1'\phi_2'} \\
    m_{\phi_1\phi_2'} & m_{\phi_2\phi_2'} & m_{\phi_1'\phi_2'} & m_{\phi_2'\phi_2'}
    \end{pmatrix}
    \begin{pmatrix}
    \phi_1\\
    \phi_2\\
    \phi_1'\\
    \phi_2'
    \end{pmatrix}
\end{equation}
where
\begin{equation}\label{Eq:masses}
    \begin{split}
    m_{\phi_1\phi_1} &\simeq {m_{11}}^2 + \frac{3}{2} v_{1}^{2} z_{1} + \frac{1}{2} w_{2}^{2} z_{12} \\
    m_{\phi_1\phi_2} &\simeq {m_{12}}^2  + \frac{3}{2} v_{1}^{2} z_{6} + \frac{1}{2} w_{2}^{2} z_{14}\\
    m_{\phi_1\phi_1'} &\simeq v_{1} w_{2} z_{13}\\
    m_{\phi_1\phi_2'} &\simeq v_{1} w_{2} z_{12}\\
    m_{\phi_2\phi_2} &\simeq {m_{22}}^2 + \frac{1}{2} v_{1}^{2}(z_{3} + z_{4} + z_{5}) + \frac{1}{2} w_{2}^{2} z_{9}\\
\end{split}
\quad
\begin{split}
    m_{\phi_2\phi_1'} &\simeq \frac{1}{2} v_{1} w_{2} z_{10} + \frac{1}{2} v_{1} w_{2} z_{11}\\
    m_{\phi_2\phi_2'} &\simeq v_{1} w_{2} z_{14}\\
    m_{\phi_1'\phi_1'} &\simeq {m_{11}}^2 + \frac{1}{2} v_{1}^{2} z_{8} + \frac{1}{2} w_{2}^{2}(z_{3} +  z_{4} + z_{5})\\
    m_{\phi_1'\phi_2'} &\simeq {m_{12}}^2 + \frac{1}{2} v_{1}^{2} z_{13} + \frac{3}{2} w_{2}^{2} z_{7}\\
    m_{\phi_2'\phi_2'} &\simeq {m_{22}}^2 + \frac{1}{2} v_{1}^{2} z_{12} + \frac{3}{2} w_{2}^{2} z_{2}.
\end{split}
\end{equation}
We consider the sizes of these terms given the constraint on relative parameter sizes from Eq.~\ref{Eq:smallzs}. We first note that the off-diagonal terms involving $\phi_1$ depend on couplings that we require to be small, with the exception of $z_{12}$ in the term $m_{\phi_1\phi_2'}$. In the diagonal mass term for $\phi_1$, $m_{\phi_1\phi_1}$, $z_{12}$ also controls the term's dependence on $w_2$. So, setting $z_{12}$ to be very small -- as was done in the original paper's benchmark point shown in Table. \ref{table:param1} -- ensures that $\phi_1$ is decoupled from the dark electroweak scale, and has minimal mixing with any other scalars. This then means that there is a mass eigenstate composed primarily of $\phi_1$, which we denote as $h$ and identify as the SM Higgs boson.

With $z_{12}$ small, the off-diagonal terms involving $\phi_2'$ are also relatively small, so we identify the dark Higgs boson $h'$ as the mass eigenstate composed predominantly of $\phi_2'$. The level of mixing between the remaining neutral real scalars, $\phi_2$ and $\phi_1'$, is controlled by $z_{10}$. As we noted earlier, this coupling is relatively large in the given benchmark point; this allows for the other two mass eigenstates to be heavy, as their masses depend on the dark electroweak scale $w$. As in the original paper, we denote these eigenstates as $J_1^0$ and $J_2^0$. We take a similar approach with the remaining mass eigenstates. Following the original paper, we name them $A_1^0$, $A_2^0$, $H^{\pm}$, and $H^{\pm\prime}$; they too couple to the dark electroweak scale $w$ and are thus much heavier than the visible Higgs boson $h$. This allows the low-energy scalar sector of this theory to contain solely an SM Higgs state, and is also relevant for meeting constraints from flavour-changing neutral current measurements.

\subsection{Yukawa couplings and flavour-changing neutral currents}

The Yukawa sector of this theory is given by
\begin{equation}
\begin{split}
    -\Lagr_Y &= y_{1 ij}^u(\Bar{q_L^i}u_R^j\Phi_1 + \Bar{q_R^{i\prime}}u_L^{j\prime}\Phi_1') + y_{1 ij}^d(\Bar{q_L^i}d_R^j\Tilde{\Phi}_1 + \Bar{q_R^{i\prime}}d_L^{j\prime}\Tilde{\Phi}_1') +y_{1 ij}^l(\Bar{l_L^i}e_R^j\Tilde{\Phi}_1 + \Bar{l_R^{i\prime}}e_L^{j\prime}\Tilde{\Phi}_1')\\
    &+ y_{2 ij}^u(\Bar{q_L^i}u_R^j\Phi_2 + \Bar{q_R^{i\prime}}u_L^{j\prime}\Phi_2') + y_{2 ij}^d(\Bar{q_L^i}d_R^j\Tilde{\Phi}_2 + \Bar{q_R^{i\prime}}d_L^{j\prime}\Tilde{\Phi}_2') +y_{2 ij}^l(\Bar{l_L^i}e_R^j\Tilde{\Phi}_2 + \Bar{l_R^{i\prime}}e_L^{j\prime}\Tilde{\Phi}_2') + H.c.
\end{split}
\end{equation}
where $\Tilde{\Phi} = i\tau_2\Phi^\star$. We note that the mirror symmetry enforces the Yukawa couplings of a doublet and its mirror counterpart to be equal. 

We are interested in how these couplings generate quark masses, as it is the quark mass spectrum in each sector that affects the running of $\alpha_3$ and $\alpha_3'$ and allows us to achieve different visible and dark QCD scales after the mirror symmetry is broken. So, we work in the Higgs basis of Eq.~\ref{Eq:newbasis}, where only $H_1$ and $H_1'$ gain VEVs. Then, the relevant Yukawa matrices are given by 
\begin{equation}\label{Eq:yukawa-matrices}
\begin{split}
    \Tilde{y}_1^q= V_L^q\left(\frac{v_1y_1^q + v_2y_2^q}{v}\right)V_R^{q\dagger},
    \quad
    &\Tilde{y}_2^q= V_L^q\left(\frac{-v_2y_1^q + v_1y_2^q}{v}\right)V_R^{q\dagger},\\
    \Tilde{y}_1^{q\prime}= W_L^q\left(\frac{w_1y_1^q + w_2y_2^q}{w}\right)W_R^{q\dagger},
    \quad
    &\Tilde{y}_2^{q\prime}= W_L^q\left(\frac{-w_2y_1^q + w_1y_2^q}{w}\right)W_R^{q\dagger},
\end{split}
\end{equation}
where $q = u,d$, and $\Tilde{y}_i^{q(\prime)}$ is the Yukawa matrix for couplings between $H_i^{(\prime)}$ and either up- or down-type quarks. $V_{L,R}^q$ and $W_{L,R}^q$ are the left- and right-handed matrices that respectively diagonalise $\Tilde{y}_1^{q}$ and $\Tilde{y}_1^{q\prime}$. 

The Yukawa matrices relevant for generating quark masses in each sector are $\Tilde{y}_1^{q}$ and $\Tilde{y}_1^{q\prime}$. In the ASB limit of Eq.~\ref{Eq:asb_limit}, we see that the visible and dark quark masses depend primarily on $y_1^q$ and $y_2^q$, respectively. This is just the statement that $\Phi_1$ and $\Phi_2'$ are the doublets primarily responsible for mass generation in their respective sectors, allowing the quark mass spectrum in each sector to be largely independent.

The secondary Yukawa matrices in each sector are not diagonal. This leads to flavour-changing neutral currents at tree-level, which are strongly suppressed in the SM and are subject to strict experimental constraints \cite{Anderson:2001nt}. This is often controlled in 2HDMs by introducing additional discrete symmetries to restrict which types of quarks each doublet can couple to; in effect, this equates to setting some of $y_1^q$ and $y_2^q$ to zero. However, in our case all of these matrices are relevant for mass generation, and must be nonzero. Thus, in the visible sector $\Phi_1$ and $\Phi_2$ will both couple to all quarks; this corresponds to a type III 2HDM, in which tree-level FCNCs are present, and must be sufficiently suppressed.

The original paper quoted an approximate result from \cite{Branco:2011iw}, where FCNC bounds were avoided in a Type III 2HDM for $m_{H_2} \gtrsim 150$ TeV. However, this bound was obtained under the assumptions that all Yukawa couplings of $H_2$ were the size of the SM top quark coupling. The more realistic Yukawa coupling selection in this model leads to much less stringent bounds \cite{Cheng:1987rs, Atwood:1996vj}; we follow the guide of the original paper, in which all stated mass values for the additional scalars are heavy enough to sufficiently suppress FCNCs. We then ensure that we consider parameter points where all additional scalars have masses at least as large as those given in Table 1 of Ref.~\cite{Lonsdale:2018xwd}.

\section{Dark Electroweak Phase Transition}\label{sec:EWPT}
To address the apparent coincidence of cosmological mass densities of Eq.~\ref{eq:Omega-coincidence}, a comprehensive dark matter theory must explain why both the particle masses and number densities of visible and dark matter are similar. As outlined in the previous section, the asymmetric symmetry breaking structure of the mirror two Higgs doublet model allows for a dark neutron-like particle with a mass a factor of a few larger than the visible proton. With the particle masses thus linked, we now need to produce related number densities $n_{\VM}$ and $n_{\DM}$.

In this section we implement electroweak baryogenesis (EWBG) as the asymmetry generation mechanism \cite{Morrissey:2012db}. EWBG occurs at a first-order electroweak phase transition (EWPT), where the transition proceeds by bubble nucleation. The Sakharov conditions \cite{Sakharov:1967dj} are satisfied by out-of-equilibrium $C$- and $CP$-violating Yukawa interactions at the bubble walls together with $B$-violating electroweak sphaleron processes, and thus a baryon asymmetry is generated during the transition.

While all these ingredients are present within the SM, the visible electroweak phase transition (vEWPT) is crossover, not first-order \cite{Gurtler:1997kh}. Even if that was not so, the $CP$-violation in the SM Yukawa matrix would be insufficient to generate the required asymmetry \cite{Gavela:1993ts}. In our model, however, there will be a dark electroweak phase transition (dEWPT) in which the $\Phi_2'$ gains a VEV of order $w$. Its dynamics are controlled by the scalar and Yukawa couplings of the second Higgs doublet, which are only very weakly constrained by SM measurements. So, we should have the flexibility to successfully implement EWBG at the dEWPT, thus generating an asymmetry in the dark baryon number $B'$ and/or dark lepton number $L'$.

In this section we analyse the dynamics of the dEWPT, searching for parameter selections for the scalar potential $\Vfull$ of Eq. \ref{Eq:Vfull} such that we obtain a first-order electroweak phase transition that could allow for EWBG in the dark sector. We begin by constructing the finite temperature effective potential, and then specify the method by which we search for valid dark phase transitions. We find that for a number of regions of parameter space, a viable first-order EWPT can be readily achieved in the dark sector.

\subsection{The finite temperature effective potential}
We begin by constructing the finite temperature effective potential (FTEP) \cite{Dolan:1973qd, Jackiw:1974cv}, our perturbative tool for analysing the dEWPT. 

The one-loop effective potential is calculated in terms of a constant background classical field $\varphi$, and is given in general by
\begin{equation}
    V_{\mathrm{eff}}(\varphi, T) = V_{0}(\varphi) + V_1(\varphi,0) + \Delta V_1(\varphi, T),
\end{equation}
where the zero-loop contribution $V_0(\varphi)$ is just the classical tree-level potential and the one-loop contributions are split into zero-temperature and finite temperature corrections.

For our mirror two-Higgs-doublet model, the FTEP will be a function of four variables -- $\varphi_1$, $\varphi_2$, $\varphi_1'$, and $\varphi_2'$ -- as we require a real constant classical background field for each field that gains a VEV. We define the shorthand notation $f(\varphi) \equiv f(\varphi_1, \varphi_2, \varphi_1', \varphi_2')$ for any function $f$. The background fields are incorporated by defining
\begin{equation} \label{Eq:bgfielddefs}
\begin{split}
    \Phi_1 = 
    \begin{bmatrix}
    G_1^+\\
    \frac{1}{\sqrt{2}}(\varphi_1 + \phi_1 + iG_1)
    \end{bmatrix},&
    \quad
    \Phi_1' = 
    \begin{bmatrix}
    I_1^+ \\
    \frac{1}{\sqrt{2}}(\varphi_1' + \phi_1' + ia_1)
    \end{bmatrix},
    \\
    \Phi_2 = 
    \begin{bmatrix}
    I_2^+ \\
    \frac{1}{\sqrt{2}}(\varphi_2 + \phi_2 + ia_2)
    \end{bmatrix},&
    \quad
    \Phi_2' = 
    \begin{bmatrix}
    {G_2^+}'\\
    \frac{1}{\sqrt{2}}(\varphi_2' + {\phi_2}' + iG_2)
    \end{bmatrix}.
\end{split}
\end{equation}

Expanding $\Vfull$ using the above definitions and assuming real parameters, the tree-level component of the FTEP is given by
\begin{equation}
\begin{split}
    V_0(\varphi) &= \frac{1}{2}m_{11}^2\left({\varphi_1}^2 + {\varphi_1'}^2\right) + \frac{1}{2}m_{22}^2\left({\varphi_2}^2 + {\varphi_2'}^2\right) + m_{12}^2\left(\varphi_1\varphi_2 + \varphi_1'\varphi_2'\right) \\
    &+ \frac{1}{8}z_1\left({\varphi_1}^4 + {\varphi_1'}^4\right) + \frac{1}{8}z_2\left({\varphi_2}^4 + {\varphi_2'}^4\right) + \frac{1}{4}\left(z_3 + z_4 + z_5\right)\left({\varphi_1}^2{\varphi_2}^2 + {\varphi_1'}^2{\varphi_2'}^2\right) \\
    &+ \frac{1}{2}z_6\left({\varphi_1}^3\varphi_2 + {\varphi_1'}^3\varphi_2'\right) + \frac{1}{2}z_7\left(\varphi_1{\varphi_2}^3 + \varphi_1'{\varphi_2'}^3\right) + \frac{1}{4}z_8{\varphi_1}^2{\varphi_1'}^2 + \frac{1}{4}z_9{\varphi_2}^2{\varphi_2'}^2 \\
    &+ \frac{1}{2}\left(z_{10} + z_{11}\right)\varphi_1\varphi_2\varphi_1'\varphi_2' + \frac{1}{4}z_{12}\left({\varphi_1}^2{\varphi_2'}^2 + {\varphi_1'}^2{\varphi_2}^2\right) \\
    &+ \frac{1}{2}z_{13}\left({\varphi_1}^2\varphi_1'\varphi_2' + {\varphi_1'}^2\varphi_1\varphi_2\right) + \frac{1}{2}z_{14}\left({\varphi_2}^2\varphi_1'\varphi_2' + {\varphi_2'}^2\varphi_1\varphi_2\right).
\end{split}
\end{equation}

\subsubsection{One-loop corrections and renormalisation}

The one-loop corrections $V_1(\varphi, 0)$ and $\Delta V_1(\varphi, T)$ are calculated using the Coleman-Weinberg \cite{Coleman:1973jx} and finite temperature \cite{Quiros:1999jp} methods respectively, and are given by 
\begin{equation}
\begin{split}
    V_1(\varphi, 0) &= \sum_i\pm\frac{n_i}{2}\int\frac{d^4p}{(2\pi)^4}\log\left(p^2+m_i^2(\varphi)\right),\\
    \Delta V_1(\varphi, T) &= \frac{T^4}{2\pi^2}\sum_{i}\pm n_iJ_{\pm}\left(\frac{{m_i}^2(\varphi)}{T^2}\right),
\end{split}
\end{equation}
where $i$ counts over all particle species with $+$ for bosons and $-$ for fermions, and $n_i$ and $m_i(\varphi)$ are the multiplicity and field-dependent mass of species $i$. The thermal functions $J_{\pm}(y^2)$ are given by
\begin{equation}
    J_{\pm}(y^2) \equiv \int_0^\infty dx \; x^2 \log\left(1 \mp e^{-\sqrt{x^2+y^2}}\right),
\end{equation}
and will be calculated numerically via the package \texttt{CosmoTransitions} \cite{Wainwright:2011kj}.

Before specifying a renormalisation scheme for the UV-divergent integral $V_1(\varphi, 0)$, we calculate the field-dependent masses ${m_i}^2(\varphi)$. The scalar boson mass matrix is obtained from $\Vfull$ given the field definitions from Eq.~\ref{Eq:bgfielddefs}. The Goldstone bosons that are massless at the tree-level minimum will not be massless in general when we expand around the background fields, and thus must be included in the one-loop corrections. In total, 16 scalars are included: the six neutral mass eigenstates $h$, $h'$, $A_1^0$, $A_2^0$, $J_1^0$, and $J_2^0$, four charged scalars $H^{\pm}$ and ${H^{\pm}}'$, and six Goldstone bosons $G^0$, $G^{\pm}$, ${G^0}'$, and ${G^{\pm}}'$, each with a multiplicity of one.

In the visible sector the gauge boson mass matrix with respect to the basis $W^{\pm}$, $W^3$, $B$ is \cite{Cline:2006ts}
\begin{equation}
    \mathcal{M}_{\mathrm{gauge}} = \frac{{\varphi_1}^2 + {\varphi_2}^2}{4}
    \begin{pmatrix}
    g^2 & 0 & 0 & 0 \\
    0 & g^2 & 0 & 0 \\
    0 & 0 & g^2 & gg' \\
    0 & 0 & gg' & {g'}^2
    \end{pmatrix}
\end{equation}

By mirror symmetry, the dark gauge bosons have an equivalent mass matrix with respect to the basis ${W^{\pm}}'$, ${W^3}'$, $B'$, with $\varphi_1'$ and $\varphi_2'$ replacing $\varphi_1$ and $\varphi_2$. All gauge bosons have multiplicity three, corresponding to one longitudinal and two transverse modes.

The mass of the top quark dominates the contributions from the visible fermions, and is given by
\begin{equation}\label{Eq:topmass}
    {m_t}^2(\varphi) = \frac{1}{2}(y_1^t{\varphi_1}^2 + y_2^t{\varphi_2}^2),
\end{equation}
where $y_i^t$ indicates the Yukawa coupling of the doublet $\Phi_i$ to the top quark. In the dark sector, the heaviest quark is not necessarily the mirror partner of the visible top quark. We denote the Yukawa couplings of the heaviest dark quark by $y_i^h$, and obtain its mass by replacing $y_i^t$, $\varphi_1$ and $\varphi_2$ with $y_i^h$, $\varphi_1'$ and $\varphi_2'$ respectively in the above equation. Since only $\Phi_2'$ gains a VEV during the dEWPT, $y_2^h$ is the relevant parameter when considering the fermionic contributions to the FTEP; its value depends on the choice of dark Yukawa couplings. The multiplicities of the top quark and heaviest dark quark are both 12. 

With the field-dependent masses determined, the next point to address is the choice of renormalisation scheme for the zero-temperature one-loop corrections. We will use a cutoff regularisation scheme\footnote{This scheme differs from the standard $\overline{MS}$ dimensional regularisation commonly applied in two-Higgs-doublet models. This is due to the disparate electroweak scales in our model, $v$ and $w$, which lead to the FTEP being highly sensitive to the choice of renormalisation scale $\mu$. In particular, the tree-level minimum will change drastically for differing values of $\mu$, whereas cutoff regularisation automatically preserves tree-level minima.} given by\begin{equation}
    V_1(\varphi,0) = \sum_i \pm \frac{n_i}{64\pi^2} \left[{m_i}^4(\varphi) \left(\log\frac{{m_i}^2(\varphi)}{{m_i}^2(v)} - \frac{3}{2} \right) + {m_i}^2(\varphi){m_i}^2(v)\right],
\end{equation}
where ${m_i}^2(v)$ indicates that the mass is calculated at the tree-level minimum given by $\varphi_i = v_i$ and $\varphi_i' = w_i$. 

However, $\log({m_i}^2(\varphi)/{m_i}^2(v))$ is logarithmically divergent for $i = G^0$, $G^{\pm}$, ${G^0}'$, and ${G^{\pm}}$, as the Goldstone bosons are massless at the tree-level minimum, . This has been addressed in both the standard model \cite{Espinosa:1992kf} and a two-Higgs-doublet model \cite{Cline:2011mm}; the issue is alleviated by adjusting the Goldstone contributions to be
\begin{equation}
    \sum_{i = G^0, G^{\pm}}\frac{n_i}{64\pi^2} \left[{m_i}^4(\varphi) \left(\log\frac{{m_i}^2(\varphi)}{{m_{\mathrm{IR}}}^2(v)} - \frac{3}{2} \right)\right],
\end{equation}
where ${m_{\mathrm{IR}}}^2(v)$ is some infrared mass scale that both references take to be ${m_h}^2(v)$, the mass of the SM Higgs boson. We adapt this condition to our situation by choosing ${m_{\mathrm{IR}}}^2(v)$ to be ${m_h}^2(v)$ for the visible Goldstone bosons ${G^0}$ and ${G^{\pm}}$, and ${m_h'}^2(v)$ for the dark Goldstone bosons ${G^0}'$ and ${G^{\pm}}'$.

So, altogether, the one-loop corrections are given at zero temperature by
\begin{equation}
\begin{split}
    V_1(\varphi,0) &= \sum_{i \in \mathcal{F}} \pm \frac{n_i}{64\pi^2} \left[{m_i}^4(\varphi) \left(\log\frac{{m_i}^2(\varphi)}{{m_i}^2(v)} - \frac{3}{2} \right) + {m_i}^2(\varphi){m_i}^2(v)\right] \\
    &+ \sum_{i = G^0, G^{\pm}}\frac{n_i}{64\pi^2} \left[{m_i}^4(\varphi) \left(\log\frac{{m_i}^2(\varphi)}{{m_h}^2(v)} - \frac{3}{2} \right)\right] \\
    &+ \sum_{i = {G^0}', {G^{\pm}}'}\frac{n_i}{64\pi^2} \left[{m_i}^4(\varphi) \left(\log\frac{{m_i}^2(\varphi)}{{m_h'}^2(v)} - \frac{3}{2} \right)\right],
\end{split}
\end{equation}
where $\mathcal{F} = [h, h', A_1^0, A_2^0, J_1^0, J_2^0, H^{\pm}, {H^{\pm}}', W^{\pm}, Z, {W^{\pm}}', Z', t, t']$ lists all species except the Goldstone bosons.

\subsubsection{Thermal masses}

To be able to trust our perturbative calculations at the critical temperature, we need to apply a daisy resummation procedure to account for higher-loop corrections. Considering the one-loop corrections as a function of the field-dependent masses,
\begin{equation}
    V_1({m_i}^2(\varphi), T) \equiv V_1(\varphi,0) + \Delta V_1(\varphi, T),
\end{equation}
the standard resummation methods according to Parwani~\cite{Parwani:1991gq} and Arnold and Espinosa~\cite{Arnold:1992rz}) produce
\begin{equation}
    V_{1, \mathrm{P}} = V_1({m_i}^2(\varphi, T), T)
\end{equation}
and
\begin{equation}
    V_{1, \mathrm{A-E}} = V_1({m_i}^2(\varphi), T) + \frac{T}{12\pi}\sum_{i=\mathrm{bosons}}n_i\left[{m_i}^3(\varphi) - {m_i}^3(\varphi,T)\right]
\end{equation}
respectively.

These expressions require the calculation of the thermal masses ${m_i}^2(\varphi, T)$. Only scalars and the longitudinal components of the gauge bosons gain thermal masses. These are calculated by adding thermal correction matrices to the scalar and gauge boson mass matrices prior to diagonalisation \cite{Kainulainen:2019kyp}. For the scalars, this is given by
\begin{equation}
    (\delta \mathcal{M}_{\mathrm{scalar}})_{ij} = \frac{T^2}{24}\sum_i c_i n_i\frac{\partial^2{m_i}^2(\varphi)}{\partial\varphi_i\partial\varphi_j}
\end{equation}
where $c_i$ is $1$ for bosons and $-1/2$ for fermions, and $\varphi_i$ runs over all four background fields $\varphi_1$, $\varphi_2$, $\varphi_1'$, and $\varphi_2'$. For the gauge bosons, the thermal correction matrix is
\begin{equation}
    \delta \mathcal{M}_{\mathrm{gauge}} = 2T^2
    \begin{pmatrix}
    g^2 & 0 & 0 & 0 \\
    0 & g^2 & 0 & 0 \\
    0 & 0 & g^2 & 0 \\
    0 & 0 & 0 & {g'}^2
    \end{pmatrix}.
\end{equation}

\subsection{Characterising the phase transition}
With the effective potential in hand, we can now determine the properties of the dEWPT. To allow for electroweak baryogenesis, we search for phase transitions that are first-order -- that is, where the effective potential develops multiple minima separated by energy barriers. Such a transition is characterised by the critical temperature $T_C$, determined by
\begin{equation}
    V_{\mathrm{eff}}(0)|_{T=T_C} = V_{\mathrm{eff}}(\phi_C)|_{T=T_C},
\end{equation}
where $\phi_C$ indicates the field values at the symmetry-breaking minimum at $T=T_C$. The other relevant parameter of the phase transition is its \emph{strength}, given by $\xi = \phi_C/T_C$. For the dark baryon asymmetry to avoid being washed out following the dEWPT, the sphaleron rate must be sufficiently suppressed; this corresponds to the condition $\xi \gtrsim 1$, or that the phase transition is \emph{strongly first-order}\footnote{While this is the conventional criterion used to avoid sphaleron washout in electroweak baryogenesis theories, it is not gauge invariant; both $\phi_C$ and $T_C$ suffer gauge dependence when calculated to a finite order of perturbation theory \cite{Patel:2011th, Croon:2020cgk}. Despite this, it remains in use in the literature, including in 2HDM implementations of EWBG (see e.g. \cite{Cline:1996mga, Fromme:2006cm}).}.

In this section we describe our method for calculating these properties and identifying strong first-order phase transitions. A number of difficulties arise in this discussion that produce non-negligible theoretical uncertainties in our calculations; however, we find that these are manageable, as we are not aiming to precisely calculate the strength of phase transition, but are merely ensuring that the bound $\xi \gtrsim 1$ is satisfied.

\subsubsection{Finding the critical temperature}

The actual temperature at which a first-order EWPT commences is the nucleation temperature $T_N$, where tunnelling between the minima occurs at a sufficient rate for bubbles of broken phase to be nucleated. This is a difficult quantity to precisely determine, and requires calculation of tunnelling probabilities and bubble profiles that are beyond the scope of this analysis \cite{Dine:1992wr}. $T_C$ is much easier to determine, and is a common stand-in when analysing first-order phase transitions, as it usually lies just above $T_N$. 

We now describe an algorithm for finding $T_C$. At $T = 0$, the global minimum of the effective potential is given by the ASB limit $\varphi_i = v_i$, $\varphi_i' = w_i$ where $v_1 \gg v_2$, $w_2 \gg w_1$, and $w_2 \gg v_1$. At very high temperatures, the only minimum of the effective potential is at the origin. To identify a dark first-order transition, we track the value of $\varphi_2'$ at the minimum of $V_{\mathrm{eff}}(\varphi, T)$, denoting this quantity as $w_2(T)$. As $T$ increases, we observe that $w_2(T)$ decreases from $w_2$ at $T = 0$ until a given temperature at which it has a sudden discontinuity and drops to the symmetric phase where $w_2(T) \sim 0$. The temperature at which this drop occurs is the critical temperature $T_C$. We take the transition strength to be
\begin{equation}
    \xi = w_2(T_C)/T_C,
\end{equation}
as all other background field values are much smaller than $\varphi_2'$ at the asymmetric minimum near the dEWPT.

To find the temperature at which the drop occurs, we start at $T = 0$ and begin by increasing the temperature with a large step size (on the order of $w_2$). At each new temperature $T$, we calculate $w_2(T)$ by numerically finding the minimum of the effective potential using the methods provided by the coding package \texttt{CosmoTransitions}. When $w_2(T)$ jumps to being very small (we use the condition that $w_2(T) < 1$ GeV), the symmetric minimum is now the global minimum. We then decrease the step size by an order of magnitude and begin decreasing the temperature until the asymmetric minimum becomes the global minimum again and $w_2(T)$ jumps back up to a large value. Repeating this process, we zero in with increasing accuracy on the temperature at which there is a discontinuity in $w_2(T)$, and we terminate the process when the step size is small enough that the critical temperature has been determined to a desired precision.

\begin{figure}
    \centering
    \includegraphics[]{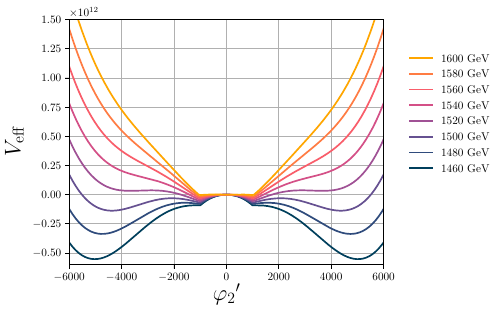}
    \caption{The effective potential $V_{\mathrm{eff}}$ as a function of $\varphi_2'$ for a range of temperatures around the critical temperature at $T_C \sim 1500$ GeV. The other background fields have been set to $\varphi_1 = \varphi_2 = \varphi_1' = 0$. The potentials at different temperatures have been vertically translated such that they coincide at the origin in order to more clearly illustrate the anomalous behaviour of $V_{\mathrm{eff}}$ for small values of $\varphi_2'$.}
    \label{fig:wiggle}
\end{figure}

However, when we apply this algorithm we run into an issue: for many parameter values of interest, $w_2(T)$ undergoes not one but two discontinuities as the temperature changes. This corresponds to the presence of a second asymmetric minimum between the symmetric minimum and the main asymmetric minimum in regions near the critical temperature. This extra minimum can be seen in Fig.~\ref{fig:wiggle}, where we plot the effective potential as a function of $\varphi_2'$ for a range of temperatures around the critical temperature, setting $\varphi_1 = \varphi_2 = \varphi_1' = 0$. This secondary minimum was identified in Ref.~\cite{Cline:1996mga} as an anomalous effect due to the presence of small or negative field-dependent particle masses. To explain and account for the presence of this artifact, we must address the perturbative validity of our daisy-resummed effective potential.

\subsubsection{The perturbative validity of the effective potential}

In Ref.~\cite{Quiros:1999jp}, the perturbative validity of the daisy resummation scheme was discussed in the context of a simple model involving a scalar singlet $\phi$ with a quartic coupling $\lambda$. They identified an expansion parameter $\lambda T/m(\phi)$ that must be small for daisy resummation to be valid. Since $m(\phi) \propto |\phi|$, the expansion parameter will be larger for smaller values of $\phi$, and indeed it is at low values of the background field that the anomalous minimum appears -- we suggest that its presence implies that our perturbative calculations are not valid when background field values are too small.

To quantify this, we need to adapt the expansion parameter for the simple case to our more complex theory. This was done in a two-Higgs-doublet model in Ref.~\cite{Cline:1996mga}; their expansion parameter was of the same form as for the simple case, with $\lambda$ chosen to be largest quartic coupling in their potential, and $m(\phi)$ taken to be the mass of the lightest of the additional scalars in the Higgs sector. Their argument for considering only the masses of the new scalars was that these provided the dominant corrections to the one-loop potential. In our case, we have a similar situation, where the heavy additional scalars -- $A_1^0$, $A_2^0$, $J_1^0$, $J_2^0$, $H^{\pm}$, and ${H^{\pm}}'$ -- give the strongest contributions to the effective potential. So, by analogy, we define the perturbative expansion parameter $\epsilon$ for our model to be
\begin{equation}\label{Eq:epsilon}
    \epsilon \equiv \frac{\max(z_i)T}{\min(m_j(\varphi))},
\end{equation}
where $z_i$ is any of the fourteen quartic couplings in $\Vfull$, and $j$ counts over the heavy scalar species listed above.

So, for a given value of $T$, and with $\varphi_1 = \varphi_2 = \varphi_1'$ = 0, there is a specific value of $|\varphi_2'|$ at which $\epsilon = 1$, which we call the \textit{perturbative boundary}. For values of $|\varphi_2'|$ below this, the expansion parameter will be greater than one and we can not trust the perturbative techniques used to calculate $V_{\mathrm{eff}}(\varphi, T)$. To account for this, we adjust the algorithm to find the temperature at which $w_2(T)$ drops discontinuously not to a value close to zero, but to a value below this perturbative boundary. This gives only an estimate for $T_C$, as we cannot measure the true critical temperature directly if we do not trust the effective potential at the origin. However, as noted before, we are not looking to calculate the specific value of the strength; as we are only identifying phase transitions satisfying the condition $\xi \gtrsim 1$, our new method for calculating $T_C$ is sufficiently accurate for our purposes.

\subsection{Results}\label{sec:ewpt-results}
We begin our search for a strong dark electroweak phase transition at a parameter point given by Table~\ref{table:resultsparam1}. This corresponds to the parameter selection from the original paper that we considered in Sec.~\ref{sec:model}, but with $z_1$ and $z_2$ each increased by a factor of two to account for an erratum in Ref.~\cite{Lonsdale:2018xwd}. These values provide a good starting point as they satisfy all conditions from Sec.~\ref{sec:model}, producing an asymmetric symmetry breaking minimum with $v = 246$ GeV and $w = 7276$ GeV. We also note that we set $y_2^h = 1$ and do not change this throughout the search. This is due to considerations from Section~\ref{sec:asym}, where we require $\mathcal{O}(1)$ dark Yukawa couplings to generate a sufficient asymmetry.

The phase transition at this parameter point is second-order. In the following sections, we alter these parameters to find regions of parameter space in which the dark electroweak phase transition is strongly first-order. We identify a number of qualitatively different regions for which this is possible, guided by considerations from Sec.~\ref{sec:asym} in which extra requirements are placed on the VEVs and particle masses of the scalar sector to ensure the feasibility of certain portal interactions.

\begin{table}
\centering
\begin{tabular}{C{5cm}C{5cm}} 
\toprule
\textbf{Parameters} & \textbf{Values}     \\ 
\hline
${m_{11}}^2$ & $-(87~\mathrm{GeV})^2$    \\
${m_{12}}^2$ & $-(90~\mathrm{GeV})^2$     \\
${m_{22}}^2$ & $-(2600~\mathrm{GeV})^2$    \\
$z_1$, $z_2$      & $0.258$      \\
$z_3$, $z_8$, $z_9$, $z_{10}$      & $0.8$        \\
$z_4$, $z_5$, $z_6$, $z_7$, $z_{11}$, $z_{13}$, $z_{14}$      & $0.01$       \\
$z_{12}$      & $1\times10^{-8}$  \\
\bottomrule
\end{tabular}
\caption{The initial parameter selection for the scalar potential $\Vfull$ at which we start the search of the parameter space.}\label{table:resultsparam1}
\end{table}

\subsubsection{Decreasing the mass of the dark Higgs boson}\label{sec:h'}

\begin{figure}
    \centering
    \includegraphics[]{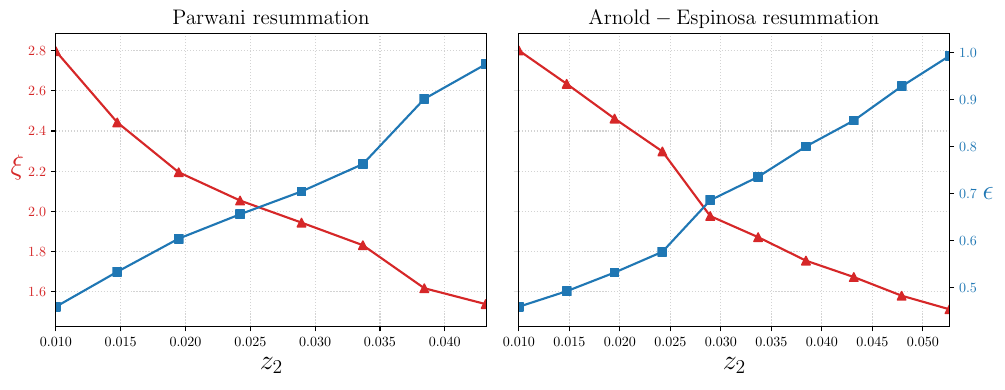}
    \caption{Plots of the phase transition strength $\xi$ (triangles) and perturbative expansion parameter $\epsilon$ (squares) at the critical temperature for different values of the quartic coupling $z_2$. The left-hand plot uses the Parwani daisy resummation scheme, and right-hand plot uses the Arnold-Espinosa resummation scheme. These values are calculated for $\rho = 30$ and $y_2^h = 1$. Other unspecified parameters of the scalar potential are as described in the text.}
    \label{fig:z2smallp}
\end{figure}

In the standard model the EWPT is expected to be first-order for small values of the Higgs mass \cite{Gurtler:1997kh}. So, we adjust our parameters such that the dark Higgs mass $m_h'$ is reduced: the relevant coupling to consider is $z_2$. To keep $w$ the same when $z_2$ is altered, we set ${m_{22}}^2 = -\frac{1}{2}z_2w^2$. In Fig.~\ref{fig:z2smallp}, we plot the strength of the phase transition, $\xi$, in red for small values of $z_2$. As expected, for low values of $z_2$ the dEWPT is sufficiently strong, and gets stronger as $z_2$ gets smaller. This general behaviour remains true for both and the Parwani and Arnold-Espinosa resummation procedures -- given by the left-hand and right-hand plots, respectively -- with the latter method producing slightly stronger transitions for the same value of $z_2$. In the same figure we also plot the value of the perturbative expansion parameter $\epsilon$ at the critical temperature in blue. Since we can only trust our calculation of $V_{\mathrm{eff}}(\varphi, T)$ for $\epsilon < 1$, we only extend the graph until the value of $z_2$ at which $\epsilon$ becomes too large. The inverse relationship between $\epsilon$ and $\xi$ implies that our perturbative methods are most reliable when the phase transition is very strong, which allows us to be confident in our results in the relevant regions of parameter space.

\subsubsection{Increasing the dark electroweak scale}\label{sec:rho}

\begin{figure}
    \centering
    \includegraphics[]{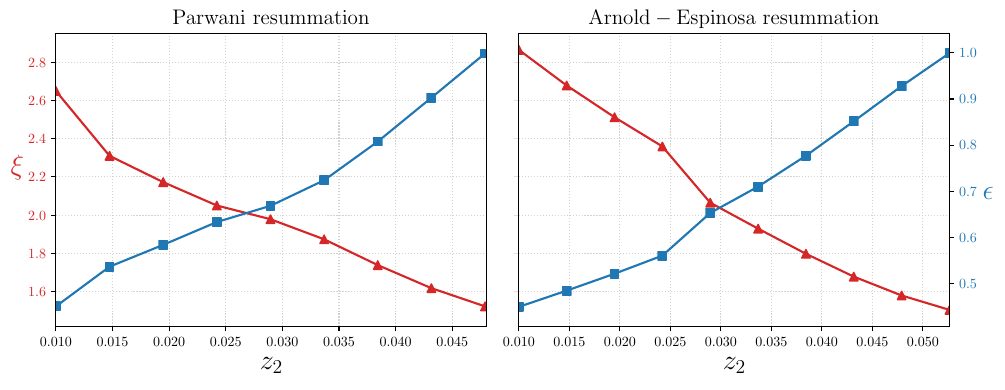}
    \caption{Plots of the phase transition strength $\xi$ (triangles) and perturbative expansion parameter $\epsilon$ (squares) at the critical temperature for different values of the quartic coupling $z_2$. The left-hand plot uses the Parwani daisy resummation scheme, and right-hand plot uses the Arnold-Espinosa resummation scheme. These values are calculated for $\rho = 1000$ and $y_2^h = 1$. Other unspecified parameters of the scalar potential are as described in the text.}
    \label{fig:z2largep}
\end{figure}

We now explore what happens when we alter the dark electroweak scale $w$. This is motivated by the upcoming discussion in Sec.~\ref{sec:asym}, where varying the dark electroweak scale gives us greater freedom in choosing portal interactions. From now on we follow the lead of the original paper and work primarily with the quantity $\rho \equiv w/v$. Since the value of $v$ is fixed by the VEV of the SM Higgs boson, we have that $w$ = $(246\rho)$ GeV.

When we increase $\rho$, the only necessary change to make to the parameters of Table~\ref{table:resultsparam1} to satisfy the constraints of Sec.~\ref{sec:model} is to decrease $z_{14}$. For $\rho = 1000$, we set $z_{14} = 1 \times 10^{-4}$. As before, we calculate the strength and expansion parameter of the phase transition for small $z_2$; these are shown in Fig.~\ref{fig:z2largep}, and are qualitatively very similar to the results of Fig.~\ref{fig:z2smallp}, where $\rho = 30$. So, for a given selection of quartic couplings, altering $\rho$ has minimal effect on the strength of the phase transition.

\subsubsection{Decreasing the masses of the additional scalars}\label{sec:mH}

\begin{figure}
    \centering
    \includegraphics[]{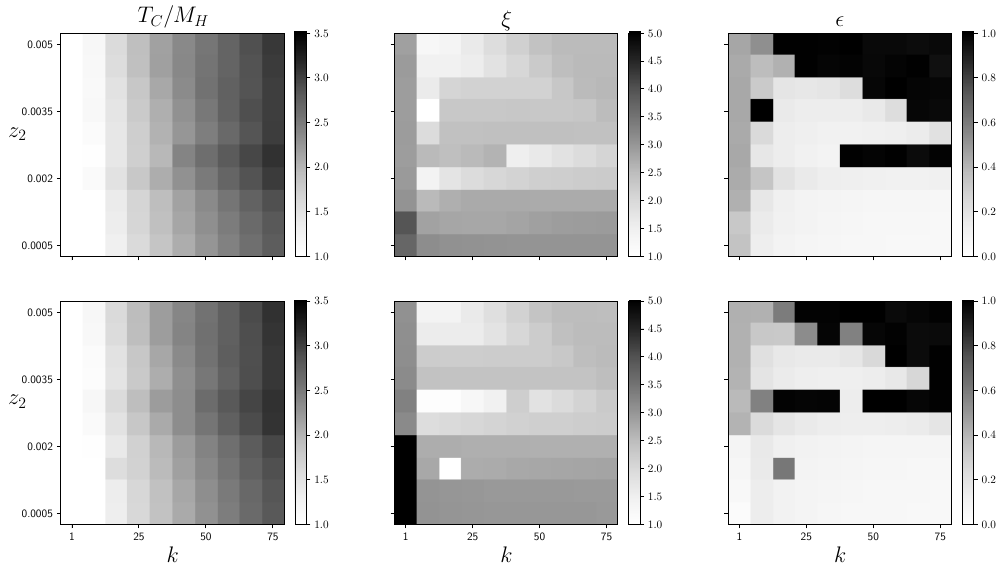}
    \caption{Plots of the ratio of the critical temperature $T_C$ and heaviest additional scalar mass $M_H$ (left-most column), phase transition strength $\xi$ (middle column), and perturbative expansion parameter $\epsilon$ at the critical temperature (right-most column) for parameter points with a varying scalar coupling $z_2$ and scaling factor $k$. The upper (lower) plot in each column corresponds to calculations using the Parwani (Arnold-Espinosa) daisy resummation scheme. These values were calculated for $\rho = 300$ and $y_2^h = 1$. Other unspecified parameters of the scalar potential are as described in the text.}
    \label{fig:TMt}
\end{figure}

In these regions of parameter space, while the small values of $z_2$ lower the mass of the dark Higgs boson $h'$, the additional scalars -- $A_1^0$, $A_2^0$, $J_1^0$, $J_2^0$, $H^{\pm}$, and ${H^{\pm}}'$ -- remain heavy. This allows for the FCNC constraints to be met; however, the lepton portal interactions in Sec.~\ref{sec:asym} require these heavy scalars to be in thermal equilibrium, and thus they must have masses lower than the critical temperature of the dark electroweak phase transition, $T_C$. This cannot be achieved with the parameter selections we have identified so far. For example, consider a strong first-order dEWPT phase transition at which $z_2 = 0.02$, $\rho = 30$, ${m_{22}}^2 = -\frac{1}{2}z_2w^2$, and all other parameters are given as in Table \ref{table:resultsparam1}. At this parameter point we have a strong-first order transition with $\xi = 2.2$ and $T_C = 1715$ GeV, while the heavy scalar masses range between $4500$ and $4800$ GeV; so, we must identify new regions of parameter space for these masses to lie under $T_C$.

The masses of the additional scalars depend primarily on $z_3$ and $z_9$, so we must reduce these parameters. However, to maintain an asymmetric symmetry breaking structure for the scalar potential, we must satisfy the constraint that $z_3, z_8, z_9 \gg z_6, z_7, z_{11}, z_{13}, z_{14}$. To roughly achieve both requirements in way that can be easily parametrised, we divide all scalar couplings $z_i$ (with $i$ counting from 3 to 14) by a scaling factor $k$. This leaves $z_1$ and $z_2$ unchanged -- $z_1$ must remain fixed to preserve the tree-level mass of the visible Higgs boson at the standard model value of 125 GeV, and we vary $z_2$ independently from $k$ to find parameter points for which the phase transition is strongly first-order and the additional scalars are sufficiently light. 

As the additional scalars are now lighter, we have a potential conflict with the FCNC constraints. We can avoid this concern by working in regions with large values of $\rho$, as the higher dark electroweak scale $w$ raises both the critical temperature and the masses of the additional scalars. This allows these masses to be large enough to suppress the FCNCs while still being smaller than the temperature of the dEWPT.

For our search we choose $\rho = 300$; this necessitates that prior to being divided by the scaling factor $k$, $z_{14}$ is set to $1 \times 10^{-3}$. The eleven scalar couplings from $z_3$ to $z_{13}$ again have their values before scaling given by those from Table \ref{table:resultsparam1}. We vary the scaling factor $k$ from 1 up to 75, and work in a region of very small values for $z_2$ given by $5 \times 10^{-4} \leq z_2 \leq 5 \times 10^{-3}$. The results of this scan are given in Fig.~\ref{fig:TMt}, where all calculations are performed using both the Parwani and Arnold-Espinosa daisy resummation schemes. The left-most column of figures gives the value of the ratio $T_C/M_H$, where $T_C$ is the critical temperature of the dEWPT and $M_H$ is the mass of the heaviest additional scalar. Regions in white have $T_C/M_H < 1$, indicating parameter selections for which the additional scalars are not lighter than the dEWPT temperature. The middle column of figures show the strength of the phase transition; here, white regions indicate that $\xi < 1$, and thus give parameters for which the phase transition is not sufficiently strong. The final column of figures give the perturbative expansion parameter $\epsilon$ at $T_C$. The black regions of these plots correspond to $\epsilon > 1$, and thus indicate regions where we cannot trust the calculations at the critical temperature. So, the valid parameter points in this region of parameter space are those that are not contained in any of these forbidden regions.

These results display some features that we expect to see, and some that are more curious. Increasing the scaling factor $k$ allows for the masses of the additional scalars to be below the critical temperature; decreasing $z_2$ both strengthens the phase transition and increases the validity of the perturbative expansion, as expected. However, there are notable artifacts in our results: points where the perturbative expansion parameter unexpectedly becomes large again for low values of $z_2$. While similar features occur when using both daisy resummation schemes, they do not occur at the same parameter points, so we take them to be anomalous effects.

\subsubsection{Summary of results}

In this section we identified a number of regions of parameter space for which the dark electroweak phase transition is strongly first-order, providing the out-of-equilibrium dynamics that allow for electroweak baryogenesis to occur in the dark sector. To summarise these results, in Table \ref{table:samplepoints} we give an example parameter selection for each region, along with a link to the subsection in which that region was motivated. To keep the results concise, we only state the parameters of $\Vfull$ that differ from those listed in Table \ref{table:resultsparam1}, noting that $m_{22}^2$ is given by $-\frac{1}{2}z_2(v\rho)^2$. The exception to this is when the scaling factor $k$ is increased. Then, the quoted values of all parameters from $z_3$ to $z_{14}$ must be reduced by this scaling factor, with the value of $z_{14}$ given prior to scaling. For each parameter point we list the strength of the phase transition as well as two parameters relevant to the discussions of the following section:  the critical temperature $T_C$ and heaviest scalar mass $M_H$.

\begin{table}
\centering
\begin{tabular}{C{2cm}|C{1.2cm}C{1.2cm}C{.8cm}C{1cm}|C{1cm}C{2cm}C{2cm}} 
\toprule\hline
\textbf{Section} & $z_2$ & $z_{14}$ & $k$ & $\rho$ & $\xi$ & $T_C$ [GeV] & $M_H$ [GeV] \\ 
\hline
\ref{sec:h'} & $0.01$ & $0.01$ & $1$ & $30$& $2.7$ & $1.5 \times 10^3$ & $4.8 \times 10^3$ \\ 
\ref{sec:rho} & $0.02$ & $0.0001$ & $1$ & $1000$& $2.6$ & $ 5.0 \times 10^4$ & $ 1.6 \times 10^5$ \\
\ref{sec:mH} & $0.0035$ & $0.001$ & $25$ & $300$ & $2.3$ & $1.7 \times 10^4$ & $8.8 \times 10^3$  \\
\hline\bottomrule
\end{tabular}
\caption{Example parameter points for each of the regions of parameter space discussed in the sections indicated in the first column. }\label{table:samplepoints}
\end{table}

\section{Asymmetry Reprocessing}\label{sec:asym}
Following electroweak baryogenesis at the dark electroweak phase transition, asymmetries are produced in the dark particle numbers $B'$ and/or $L'$. In this section we briefly discuss the generation of these asymmetries and investigate the cross-sector portal interactions through which these asymmetries may be transferred to the visible sector. Our goal is to transfer the asymmetries in such a way that we reproduce the 5:1 ratio between the present-day cosmological mass densities of visible and dark matter, $\Omega_{\DM} \simeq 5 \Omega_{\VM}$, where $\Omega_X = n_X m_X/\rho_c$. Since both visible and dark matter are predominantly comprised of stable baryonic matter, then $n_{\VM}$ and $n_{\DM}$ are proportional to the net baryon numbers $B$ and $B'$ and $m_{\VM}$ and $m_{\DM}$ are proportional to the confinement scales $\Lambda_\mathrm{QCD}$ and $\Lambda_{\DM}$. We can then recast the mass density relationship as
\begin{equation}\label{Eq:5ratio}
    \frac{B'}{B}\frac{\Lambda_{\mathrm{DM}}}{\Lambda_{\mathrm{QCD}}} \simeq 5.
\end{equation}

In the original paper \cite{Lonsdale:2018xwd}, the thermal leptogenesis mechanism produced equal baryon asymmetries between the two sectors, and so only the relative sizes of the confinement scales could reproduce the ratio. As the standard model confinement scale $\Lambda_{\mathrm{QCD}} \sim 200$ MeV, this required $\Lambda_{\DM} \sim 1$ GeV. In our work, the transfer of baryon asymmetry from the visible sector to the dark sector can produce a range of different values for the ratio $B'/B$, and thus allow for greater variance in $\Lambda_{\DM}$. However, we require that $\Lambda_{\DM} > \Lambda_{\mathrm{QCD}}$, as this is necessary to lower the temperature of the dark sector relative to the visible sector and satisfy bounds from big bang nucleosynthesis. So, the asymmetry reprocessing must be able to produce a baryon number ratio satisfying
\begin{equation}\label{Eq:Bcond}
    \frac{B'}{B} \lesssim 5.
\end{equation}

Directly after the dEWPT phase transition, the initial conditions are $B, L = 0$ and $B', L' \neq 0$, with the initial dark particle asymmetries determined by the specifics of dark EWBG. In this paper we consider dark EWBG proceeding solely through $CP$-violating Yukawa interactions involving the dark Higgs doublet $H_1'$. We do not provide a full calculation of the asymmetry generation, but show that this new source of $CP$ violation can readily reproduce a sufficiently large dark asymmetry if at least one dark Yukawa coupling is $\mathcal{O}(1)$. Since we do not investigate the specifics of the asymmetry, we leave the initial asymmetry in $B'$ and $L'$ as free parameters. Their relative sizes can then be restricted by the condition Eq.~\ref{Eq:Bcond}.

In this section we analyse the asymmetry transfer by working with chemical potentials, where for a relativistic species $i$, its chemical potential $\mu_i$ is related to its number density asymmetry (for $|\mu_i| \ll T$) by
\begin{equation}
    n_i - \Bar{n}_i = \frac{g_i T^3}{6}
    \begin{cases}
    \beta\mu_i,& (\mathrm{fermions})\\
    2\beta\mu_i,& (\mathrm{bosons})
    \end{cases}
\end{equation}\label{Eq:chem_pot_and_n}
where $g_i$ is the multiplicity of the particle species and $\beta = 1/T$ \cite{Buchmuller:2000nq}. At a given temperature, constraints on these potentials arise from the reactions that are in thermal equilibrium. If there are fewer constraints than chemical potentials, then there is a conserved charge associated with each free parameter, and we can determine the ratio $B'/B$ in terms of the initial values of the conserved charges.

We consider cross-sector effective operators that conserve a total particle number -- that is, some combination of particle numbers from each sector -- to avoid the washout of the initial dark asymmetries. There is no restriction on whether viable operators need to involve just leptonic species, just baryonic species, or both, in either sector. However, operators involving just leptonic species in the visible sector must be active prior to the vEWPT to allow electroweak sphaleron reprocessing to generate a visible baryon asymmetry. 

\subsection{\emph{CP} violation and asymmetry generation}

We begin by briefly discussing the generation of the dark baryon asymmetry through electroweak baryogenesis at the dark electroweak phase transition. In this work we take the the source of $CP$ violation to be the `dark CKM matrix',
\begin{equation}
    W_{\mathrm{CKM}} = W_L^uW_L^{d\dagger}
\end{equation}
where $W_L^u$, $W_L^d$ are the left-handed diagonalisation matrices for the dark quark mass matrix as given in Eq.~\ref{Eq:yukawa-matrices}. This mirrors the implementation of EWBG in the minimal standard model, where $CP$ violation is only present in the SM CKM matrix.

We note that this differs from the standard approach to electroweak baryogenesis in two-Higgs-doublet models, where the complex terms in the scalar potential provide the new source of $CP$ violation. In this work we take the scalar potential to be real, and so do not consider this possibility; however, it is a valid alternative that is worth exploring. Complex scalar couplings would indeed provide a strong source of $CP$ violation, while also altering the scalar mass spectrum of the theory and potentially contributing sizable electric dipole moments.

In the minimal SM implementation of electroweak baryogenesis, the $CP$ violation in the CKM matrix is insufficient to generate the required asymmetry \cite{Gavela:1993ts, Huet:1994jb}. The standard rough argument presented in the literature is that the size of $CP$ violation in the CKM matrix can be characterised by the invariant
\begin{equation}
    d_{CP} = J\left({m_t}^2-{m_u}^2\right)\left({m_t}^2-{m_c}^2\right)\left({m_c}^2-{m_u}^2\right)\left({m_b}^2-{m_d}^2\right)\left({m_b}^2-{m_s}^2\right)\left({m_s}^2-{m_d}^2\right),
\end{equation}
where $J \sim 10^{-5}$ is the Jarlskog invariant and $m_q$ are the quark masses. In order to make a comparison to the baryon asymmetry, we consider the dimensionless parameter
\begin{equation}
    \delta_{CP} = \frac{d_{CP}}{T^{12}}
\end{equation}
where $T$ is taken to be $100$ GeV, the typical temperature scale of electroweak baryogenesis. This then gives $\delta_{CP} \sim 10^{-20}$, far smaller than the baryon asymmetry $\eta \sim 10^{-10}$.\footnote{The validity of this naive argument has been questioned \cite{Farrar:1993hn}, as it treats the Yukawa interactions as perturbations at $T \sim 100$ GeV. For quarks with momentum $p \sim m_q$, this estimate will not be valid and a larger asymmetry may be realised. However, this effect was found to be largely suppressed by the damping of the quarks into the plasma after their reflection off the bubble wall \cite{Gavela:1993ts}. These more detailed analyses of $CP$ violation in the SM did indeed find very small asymmetries close to $10^{-20}$.}

The $CP$ violation in the dark CKM matrix is characterised by a similar invariant $d_{CP}'$ which we now write as
\begin{equation}
    d_{CP}' = J'\left(\frac{v\rho}{\sqrt{2}}\right)^{12}\prod_{i > j}\left(y_{u_i'}^2-y_{u_j'}^2\right)\left(y_{d_i'}^2-y_{d_j'}^2\right),
\end{equation}
where $J'$ is the Jarlskog invariant for the dark CKM matrix and $\tilde{y}_{q_i'}$ is the Yukawa coupling of the dark quark $q_i'$ with generation index $i \in \{1,2,3\}$. We then consider the dimensionless parameter
\begin{equation}
    \delta_{CP}' = \frac{d_{CP}'}{T'^{12}}
\end{equation}
where $T'$ is the relevant temperature scale for dark EWBG, which we take to be $T_C$, the critical temperature of the dEWPT.

As there are no observational constraints on the values of the dark CKM matrix, we are able to alter $J'$ and the dark Yukawa couplings. We consider $\rho$ and $T_C$ values given by the parameter points from Sec.~\ref{sec:ewpt-results}. Using the SM values for the Jarlskog invariant and Yukawa couplings, we obtain $\delta_{CP}' \sim 10^{-17}$. However, taking $J'$ to be $10^{-1}$, and increasing $y_{d_3'}$ and $y_{u_2'}$ each by a factor of 10, we then obtain $\delta_{CP}' \sim 10^{-7}$. So, given the freedom we have in the dark CKM matrix, this naive estimate suggests that we can readily achieve sufficient $CP$ violation to generate a large dark baryon asymmetry as long as at least one Yukawa coupling is $\mathcal{O}(1)$. This is the only restriction we apply as a result of this discussion; it informed our choice of parameter points in Sec.~\ref{sec:EWPT}, and will inform our discussion in the remainder of this section.

As with any new source of $CP$ violation, the main phenomenological consideration is its effect on various electric dipole moments. The $CP$ violation in the dark CKM matrix generates EDMs for the dark electron and neutron similarly to the SM contributions to the visible electron and neutron EDMs. In the SM these contributions arise at a high loop order and are many orders of magnitude smaller than the current experimental limits \cite{Pospelov:2005pr, Fukuyama:2012np}. In the dark sector, the dominant diagrams that contribute to the dark EDMs could differ slightly -- depending on the dark Yukawa coupling structure -- but we expect them to produce similar contributions. The main enhancement to the dark sector EDM contributions will come from the larger Jarlskog invariant we are assuming, which is at most an increase of around four orders of magnitude.

However, we have no constraints on the dark EDMs, and so can only consider the extra contribution they may provide to the visible EDMs. The EDMs for the dark electron and neutron will produce EDMs for their visible counterparts through diagrams involving mixing between the sectors. This will generically introduce large suppression factors: for example, any diagram involving photon-dark photon mixing will be suppressed by the kinetic mixing parameter $\epsilon$ which is constrained by experiment \cite{Badertscher:2006fm} to be smaller than $1.55\times10^{-7}$. Other forms of mixing between sectors -- say by the cross-sector portal interactions we will consider later in this section -- will provide less drastic suppression factors, but will nevertheless only suppress the visible EDM contributions, not enhance them. Additionally, to avoid kinetic mixing in these diagrams, one must introduce extra loops involving charged visible particles, further reducing these terms. Taken all together, we then argue that the additional $CP$ violation in the dark CKM matrix will only provide negligible contributions to the observable visible EDMs.

\subsection{Chemical equilibrium conditions}
Before addressing any specific cross-sector portal interactions, we list the general chemical potential constraints that hold separately in each sector. We also discuss the temperatures at which these interactions will be in thermal equilibrium.

\subsubsection{The visible sector}

For the effective operators we consider later in this section, the asymmetry transfer will occur before the vEWPT. We assign chemical potentials to the Higgs doublets $\Phi_a$, the left-handed lepton doublets $l_{iL}$, the right-handed leptons $e_{iR}$, the left-handed quark doublets $q_{iL}$, and the right-handed quarks $u_{iR}$ and $d_{iR}$\footnote{Note that each doublet only receives one chemical potential. This is due to weak interactions involving the $W^{\pm}$ gauge bosons, which are massless above the scale of electroweak symmetry breaking and thus have $\mu_{W^{\pm}} = 0$.}, where $a = 1,2$ and $i = 1,2,3$ are generation indices. We choose to work in the diagonal Yukawa basis for the quark fields as in Eq.~\ref{Eq:yukawa-matrices}; thus Cabibbo mixing between left-handed quarks of different generations implies that $\mu_{q_{1L}} = \mu_{q_{2L}} = \mu_{q_{3L}} \equiv \mu_{q_L}$\footnote{PMNS mixing would imply an equivalent relationship for left-handed leptons. However, as we do not specify a neutrino mass generation mechanism in this work, we keep the analysis general by maintaining independent chemical potentials for left-handed leptons of different generations, as in Refs.~\cite{Foot:2003jt, Foot:2004pq}.}. When both Higgs doublets are in thermal equilibrium, mixing between them sets $\mu_{\Phi_1} = \mu_{\Phi_2} \equiv \mu_{\Phi}$.

When in thermal equilibrium, Yukawa interactions provide the following restrictions \cite{Buchmuller:2000nq}:
\begin{equation}\label{Eq:visyuk}
\begin{split}
    &\qL + \mphiv - \uR = 0, \quad \qL - \mphiv - \dR = 0,\\ 
    &\lL - \mphiv - \eR = 0.
\end{split}
\end{equation}
There are also restrictions from sphaleron processes -- above the vEWPT transition, both electroweak and QCD sphalerons will be in equilibrium, leading to the conditions \cite{Kuzmin:1985mm, Mohapatra:1991bz}
\begin{equation}\label{Eq:ewsphv}
    9\mu_{q_L} + \sum_{i=1}^3\mu_{l_{iL}} = 0, \quad 6\mu_{q_L} - \sum_{i=1}^3(\uR + \dR) = 0.
\end{equation}
The final condition to consider above the EWPT is the hypercharge neutrality of the universe, which gives the relation \begin{equation}\label{Eq:hypv}
    2N_{\Phi}\mphiv + 3\mu_{q_L} + \sum_{i=1}^3(2\mu_{u_{iR}} - \mu_{d_{iR}} - \mu_{l_{iL}} - \mu_{e_{iR}}) = 0,
\end{equation}
where $N_{\Phi}$ is the number of Higgs bosons that are in thermal equilibrium. 

While hypercharge neutrality applies at all temperatures above the vEWPT, the other relations quoted above only apply when the interaction is in thermal equilibrium. Above the vEWPT temperature, both sphaleron processes are in thermal equilibrium for all temperatures $T < 10^{12}$ GeV \cite{Buchmuller:2000nq}. For a Yukawa interaction with dimensionless coupling $\lambda$, the approximate rate $\Gamma \sim \lambda^2T$ implies that a given Yukawa interaction is in equilibrium for $T \lesssim \lambda^2 10^{16}$ GeV. Lighter fermions thus enter thermal equilibrium at lower temperatures.

Finally, we give the combinations of chemical potentials that correspond to the visible baryon and lepton numbers, $B$ and $L = \sum_{i=1}^3L_i$:
\begin{equation}
\begin{split}
    B &\leftrightarrow 6\qL + \sum_{i=1}^3(\uR + \dR), \\
    L_i &\leftrightarrow 2\lL + \eR.
\end{split}
\end{equation}

\subsection{The dark sector}

In the dark sector we only consider asymmetry transfer at temperatures below the dEWPT. At this transition, the ${W^{\pm}}'$ gauge bosons will become massive and initially gain a chemical potential; thus, we assign a different chemical potential for each field in a doublet. Due to the parity symmetry between the visible and dark sectors, these doublets will be right-handed, and we define their field content by
\begin{equation}
    q_{iR}' = 
    \begin{pmatrix}
    u_{iR}' \\
    d_{iR}'
    \end{pmatrix},
    \quad
    l_{iR}' = 
    \begin{pmatrix}
    \nu_{iR}' \\
    e_{iR}'
    \end{pmatrix},
    \quad
    \Phi_{jR}' = 
    \begin{pmatrix}
    {\phi_j^+}' \\
    {\phi_j^0}'
    \end{pmatrix},
\end{equation}
where $i$ and $j$ are generation indices. We assign a chemical potential to each of these fields, as well as to the left-handed leptons $e_{iL}'$ and left-handed quarks $u_{iL}'$ and $d_{iL}'$. Here we work in the diagonal Yukawa basis for the dark quarks\footnote{In this basis the dark fields we assign chemical potentials to are not the mirror partners of the visible fields that are assigned potentials. Thus, when we refer to dark fermion generations and flavours, we are not referring to the mirror counterparts of the visible fermion generations. Instead, the ``dark top'' and ``dark bottom'' quarks, for example, refer to the most massive dark quarks with dark electric charge $+2/3$ and $-1/3$, respectively.} and thus Cabibbo mixing sets $\mu_{u_{1R}'} = \mu_{u_{2R}'} = \mu_{u_{3R}'} \equiv \uRd$ and $\mu_{d_{1R}'} = \mu_{d_{2R}'} = \mu_{d_{3R}'} \equiv \dRd$. If both Higgs doublets are in thermal equilibrium, mixing between them sets $\mu_{{\phi_1^+}'} = \mu_{{\phi_2^+}'} \equiv \mu_{{\phi^+}'}$ and $\mu_{{\phi_1^0}'} = \mu_{{\phi_2^0}'} \equiv \mu_{{\phi^0}'}$. The formation of a vacuum condensate of $\phi^0$ bosons also sets $\mpOd\:=\:0$ \cite{Harvey:1990qw}.

Below the dEWPT, all dark particles are now massive. So, at temperatures below a particle's mass, Boltzmann suppression leads the reactions which involve it to fall out of thermal equilibrium. If the particle is unstable, it then decays into other species, sending its chemical potential to zero. We thus will not consider the chemical potential of species that fall out of equilibrium. In the dark sector, this will always apply to the ${W^{\pm}}'$ gauge bosons, as for all the parameter selections in Sec.~\ref{sec:EWPT}, they will be more massive than the phase transition temperature $T_C$, and thus will swiftly fall out of thermal equilibrium below the dEWPT\footnote{Although $\mu_{{W^{\pm}}'} = 0$, the fields in each doublet do not need to have equal potentials as in the visible sector. This is because the ${W^{\pm}}'$ bosons have fallen out of thermal equilibrium and so the weak interactions that related the relevant chemical potentials have now frozen out.}. 

With independent chemical potentials for the fields in each doublet, the Yukawa interactions provide twice as many constraints:
\begin{equation}\label{Eq:darkyuk}
\begin{split}
    &\uRd + \mpOd - \uL = 0, \quad \dRd - \mpOd - \dL = 0, \quad \eRd - \mpOd - \eL = 0, \\
    &\dRd + \mppd - \uL = 0, \quad \uRd - \mppd - \dL = 0, \quad \nRd - \mppd - \eL = 0.
\end{split}
\end{equation}
Below the dEWPT the dark electroweak sphaleron process is strongly suppressed while the dark QCD sphaleron remains active down until the dark quark-hadron phase transition and provides the constraint:
\begin{equation}\label{Eq:qcdsphd}
    3(\uRd + \dRd) - \sum_{i=1}^3(\uL + \dL) = 0.
\end{equation}
Even though the ${W^{\pm}}'$ bosons are too heavy for gauge interactions with on-shell ${W^{\pm}}'$ to be in equilibrium, they can act as a mediator for four-lepton and four-quark interactions. We only consider the four-lepton interactions $e_{iR}' + \bar{\nu}_{iR}' \to e_{jR}' + \bar{\nu}_{jR}'$, as the four-quark interactions involve dark right-handed quark fields that already have equal chemical potentials between generations and thus introduce no new constraints. A reaction mediated by a massive gauge boson of mass $m$ is in thermal equilibrium for $T \gtrsim (m/100~\mathrm{GeV})^{4/3}$ MeV \cite{Kolb:1990vq}. So, for even the largest values of $m_{{W^{\pm}}'}$ that we consider, these interactions are in thermal equilibrium from at least $T = 1$ GeV. This is far below the temperature ranges we are interested in, so the four-lepton reactions provide the additional constraints:
\begin{equation}\label{Eq:fourlep}
    \eRd - \nRd - \mu_{e_{jR}'} + \mu_{\nu_{jR}'} = 0.
\end{equation}
Below the dEWPT, dark hypercharge is broken to dark electric charge $Q'$. The dark charge neutrality of the universe then enforces \cite{Harvey:1990qw}:
\begin{equation}\label{Eq:Qd}
    6\uRd - 3\dRd + 2N_{\Phi'}\mppd + \sum_{i=1}^3(2\uL - \dL - \eRd - \eL) = 0,
\end{equation}
where $N_{\Phi'}$ is the number of dark Higgs doublets in thermal equilibrium.

Lastly, we state the chemical potentials combinations that correspond to the dark baryon and lepton numbers, $B'$ and $L' = \sum_{i=1}^3L_i'$:
\begin{equation}
\begin{split}
    B' &\leftrightarrow 3(\uRd + \dRd) + \sum_{i=1}^3(\uL + \dL), \\
    L_i' &\leftrightarrow \eRd + \nRd + \eL.
\end{split}
\end{equation}

\subsection{The neutron portal}\label{sec:neutron-portal}
The neutron portal operators are dimension-9 quark interactions involving one singlet up-type and two singlet down-type quarks from each sector, for example: 
\begin{equation}\label{Eq:neutportal}
    \frac{1}{M^5}\Bar{u}\Bar{d}\Bar{d}u'd's' + h.c.
\end{equation}
where we have simplified our notation by defining $u \equiv u_{1R}$, $d \equiv d_{1R}$, $u' \equiv u_{1L}'$ $d' \equiv d_{1L}'$, and $s' \equiv d_{2L}'$. This specific neutron portal operator\footnote{We note the flavour structure of this operator, involving $u'$, $d'$, and $s'$ quarks in the dark sector. If this portal instead involved one $u'$ and two $d'$ quarks, then the dark neutron would be unstable to decay into visible neutrons. As the dark neutron comprises the dark matter in the dark sector, we cannot permit this to occur.} was already considered in the original paper \cite{Lonsdale:2018xwd} in the context of satisfying bounds on dark radiation from big bang nucleosynthesis. Satisfying these phenomenological constraints is a vitally important aspect of the original model; thus, we briefly review and summarise the mechanism by which this is achieved, and in so doing motivate the neutron portal as the asymmetry transfer mechanism.

\subsubsection{Dark radiation and thermal decoupling}

The presence of dark relativistic degrees of freedom is a generic feature of ADM models that faces strong constraints from BBN and CMB measurements. These constraints can be satisfied if the temperature of the dark sector is sufficiently less than that of the visible sector at the time of BBN; to achieve this, the original paper considered a situation where the two sectors decouple between the visible and dark confinement scales (also see Ref.~\cite{Farina:2015uea}). In this temperature region, the dark quark-hadron phase transition has taken place and dark-colour confinement reduces the number of dark degrees of freedom. This allows for a large transfer of entropy from the dark to the visible sector while they are still in thermal equilibrium, causing the dark sector to cool at a faster rate than the visible sector and leading to the necessary temperature difference between the sectors.

For this process to naturally take place, there must be interactions that maintain thermal equilibrium between the visible and dark sectors down to a decoupling temperature between the two confinement sectors, $T_{\mathrm{dec}} \sim 1$ GeV. While it is not the only suggestion given for this interaction, the neutron portal of Eq.~\ref{Eq:neutportal} is the most promising candidate proposed in the original paper; once the dark quarks become confined into hadrons, they can decay through the neutron portal into unconfined visible quarks, naturally providing the mechanism through which to transfer entropy between the sectors. This portal also allows for the apparently fine-tuned decoupling temperature $T_{\mathrm{dec}}$ to arise naturally; the large masses for the dark hadrons -- gained below the dark quark-hadron phase transition by dark QCD confinement -- cause these particles to become Boltzmann suppressed, thus decoupling the neutron portal interaction in the desired temperature range.

\subsubsection{Thermal equilibrium}

The approximate rate of the neutron portal interaction is $\Gamma \sim T^{11}/M^{10}$. Comparing this to the expansion rate $H \sim T^2/m_\mathrm{Pl}$, we find that the interaction is in thermal equilibrium for the temperature range
\begin{equation}\label{Eq:neutron-portal-temp-range}
    M > T > \left(\frac{M^{10}}{m_\mathrm{Pl}}\right)^{\frac{1}{9}}.
\end{equation}
where the upper bound is from the region of validity of the effective field theory. The lower bound only applies if all quarks involved in the portal interaction have masses below the lower bound temperature; otherwise, the heaviest quark involved in the portal (of mass $m_q$) becomes Boltzmann suppressed for $T < m_q$ and the interaction falls out of thermal equilibrium.

In this section, we perform a simplified analysis in which we only consider neutron portals active at temperatures above the vEWPT temperature. This captures some of the general properties of the asymmetry transfer that this portal can achieve. However, we also lose the strong motivation from the thermal decoupling considerations, as these require neutron portals to be active down to the quark-hadron phase transition, well below the vEWPT temperature\footnote{The high-scale neutron portal we consider here can still be possible in this model if one of the other interactions from the original paper is implemented to maintain thermal equilibrium below the vEWPT temperature. The $CP$-odd Higgs-mediated interactions, for example, could fulfill this role as they do not violate lepton or baryon number within a given sector, and so could maintain thermal equilibrium down to the decoupling temperature without affecting the asymmetry transfer.}. Analysing the asymmetry transfer down to this level of around $1$ GeV introduces a large number of difficulties, which we outline in Sec.~\ref{sec:darkrad}.

To find a neutron portal that operates in this temperature range, we either take $M$ to be large enough that $(M^{10}/m_{\mathrm{Pl}})^{1/9} > T_C$, where $T_C \sim 200$ GeV is the critical temperature of the vEWPT; or, we choose a flavour structure for the specific portal operator such that it involves at least one dark quark species with a mass greater than the vEWPT temperature. Given the freedom we have in choosing the dark quark masses, the latter case is easy to implement; for example, for $\rho \sim 50$ a dark quark with a Yukawa coupling on the order of the bottom quark coupling has a mass over $200$ GeV.

\subsubsection{Equilibrium conditions}

The additional chemical potential constraint due to the neutron portal operator of Eq.~\ref{Eq:neutportal} is given by
\begin{equation}
    \mu_u + 2\mu_d - \mu_{u'} - \mu_{d'} - \mu_{s'} = 0.
\end{equation}
Similar constraints arise from neutron portal operators with different flavour structures. We note that when all Yukawa interactions involving non-Boltzmann suppressed quarks are in thermal equilibrium, the right-handed visible singlet quarks ($u_{iR}$, $d_{iR}$) and the left-handed dark singlet quarks ($u_{iL}'$, $d_{iL}'$) have equal chemical potentials between the various generations, and so the final value for $B'/B$ does not depend on the specific flavour structure of the operator.

In the visible sector, the electron is in thermal equilibrium for $T \lesssim 10^5$ GeV, and so for temperatures below $10^5$ GeV all visible Yukawa interactions are in equilibrium in addition to the electroweak and QCD spahlerons. Assuming that the mass of the additional heavy scalars is above $10^5$ GeV, $\Phi_2$ will have fallen out of equilibrium by the temperatures of interest and so for the hypercharge neutrality condition Eq.~\ref{Eq:hypv} we set $N_{\Phi}$ = 1.

In the dark sector, the dark quarks are massive and thus have the potential to become Boltzmann suppressed in our temperature range of interest. As we have at least one $\mathcal{O}(1)$ dark Yukawa coupling, the heaviest dark quark has a mass greater than the dEWPT temperature, and so falls out of thermal equilibrium by the temperature at which the portal is active. So, all dark sector Yukawa constraints will apply, except for the Yukawa interaction involving the dark top quarks. The charge neutrality and QCD sphaleron conditions must also be altered by removing the chemical potentials associated with the top quarks -- that is, removing the $\mu_{u_{3L}'}$ terms entirely and reducing the $\mu_{u_R'}$ terms by a factor of 1/3. We similarly alter the definition of $B'$. Similarly to the visible sector, we set $N_{\Phi'}$ = 1 in the charge neutrality condition.

There are 6 unconstrained chemical potentials, corresponding to six conserved charges. Not considering the portal interaction, the conserved charges in each sector are given by $B/3 - L_i$ for the visible sector and $B'$ and $L_i'$ for the dark sector. As the portal conserves $B + B'$, the six conserved charges will then be given by:
\begin{equation}
\begin{split}
    \mathcal{L}_1 &= \frac{1}{3}(B + B') - L_1, \\ 
    \mathcal{L}_2 &= \frac{1}{3}(B + B') - L_2, \\ 
    \mathcal{L}_3 &= \frac{1}{3}(B + B') - L_3, \\
\end{split}
\quad
\begin{split}
    \mathcal{L}_4 &= L_1', \\
    \mathcal{L}_5 &= L_2', \\
    \mathcal{L}_6 &= L_3'.
\end{split}
\end{equation}

The various particle numbers in the two sectors can then be expressed as a linear combination of these conserved charges, as per
\begin{equation}\label{eq:BLs}
B = \sum_{i=1}^6a_i\mathcal{L}_i,
\quad
L = \sum_{i=1}^6b_i\mathcal{L}_i,
\quad
B' = \sum_{i=1}^6c_i\mathcal{L}_i, 
\quad
L' = \sum_{i=1}^6d_i\mathcal{L}_i. 
\end{equation}
Solving the linear system of chemical potential constraints then allows us to calculate the values of the coefficients in these expressions. As an example, we give the results of this calculation in Table \ref{table:abcd}.

\begin{table}
\centering
{\begin{tabular}{C{1cm}C{2cm}C{2cm}C{2cm}C{2cm}} 
\toprule
$i$ & $a_i$ & $b_i$ & $c_i$ & $d_i$\\
\hline
1 & $\frac{476}{1959}$ & $\frac{-289}{653}$& $\frac{158}{5877}$& $0$\\
2 & $\frac{476}{1959}$ & $\frac{-289}{653}$& $\frac{158}{5877}$& $0$\\
3 & $\frac{476}{1959}$ & $\frac{-289}{653}$& $\frac{158}{5877}$& $0$\\
4 & $\frac{-56}{5877}$ & $\frac{-34}{1959}$& $\frac{158}{5877}$& $0$\\
5 & $\frac{-56}{5877}$ & $\frac{-34}{1959}$& $\frac{158}{5877}$& $0$\\
6 & $\frac{-56}{5877}$ & $\frac{-34}{1959}$& $\frac{158}{5877}$& $1$\\
\bottomrule
\end{tabular}}
\caption{The values of the coefficients defined in Eq.~\ref{eq:BLs}.}
\label{table:abcd}
\end{table}

To obtain the final ratios of particle numbers, we simply specify the initial conditions for the particle asymmetries. For example, we may assume that we start only with an asymmetry in dark baryon number $B'$, and call this value $X$. Then, the only nonzero conserved charges are $\mathcal{L}_1 = \mathcal{L}_2 = \mathcal{L}_3 = X/3$, and we obtain
\begin{equation}\label{Eq:BLnumbers}
\begin{split}
&B = \sum_{i=1}^3\frac{a_iX}{3},
\quad
L = \sum_{i=1}^3\frac{b_iX}{3},
\\
&B' = \sum_{i=1}^3\frac{c_iX}{3},
\quad
L' = \sum_{i=1}^3\frac{d_iX}{3}.
\end{split}
\end{equation}

Although we can now directly calculate $B'/B$ from these results, that only gives the ratio of particle numbers directly after the portal falls out of thermal equilibrium. As this occurs at a temperature that is still above the visible electroweak phase transition, $B$ is violated by the visible electroweak sphaleron process. $B - L$ is preserved, however, and we can relate it to $B$ after the freeze-out of the neutron portal operation by the standard relationship \cite{Buchmuller:2000nq} 
\begin{equation}
    B = \frac{28}{79}(B - L).
\end{equation}

Below the visible electroweak phase transition, $B$ is conserved, and the ratio between the final baryon numbers $B_f'$ and $B_f$ is given by
\begin{equation}
\begin{split}
    \frac{B_f'}{B_f} &= \frac{79}{28}\frac{B'}{B - L} \\
    &= \frac{79}{28}\frac{c_1 + c_2 + c_3}{a_1 + a_2 + a _3 - b_1 - b_2 - b_3}.
\end{split}
\end{equation}

\subsubsection{Results}

The results for the neutron portal are given in Table \ref{table:results_np}. We note that the ratio satisfies the rough condition of Eq.~\ref{Eq:Bcond} that $B'/B \lesssim 5$, ensuring that the dark confinement scale lies above the visible confinement scale. From Eq.~\ref{Eq:5ratio}, and taking $\Lambda_{\mathrm{QCD}} \sim 200$ MeV, we can also calculate the dark confinement scale $\Lambda_{\DM}$. This then allows us to restrict the permissible values of $\rho$, as given in the last column of the table. This restriction is determined from Fig.~1 in the original paper \cite{Lonsdale:2018xwd}, which plots $\Lambda_{\DM}$ against $\rho$ for a selection of different choices for the spectrum of Yukawa couplings to $\Phi_2$. For the given value of $\Lambda_{\DM}$, we can only choose values of $\rho$ for which the necessary Yukawa spectrum matches the requirement that $y_2^h = 1$. We note that in this simplified analysis, the neutron portal successfully reprocess the asymmetries for smaller values of $\rho$, due to the requirements on the dark Yukawa couplings.

\begin{table}
\centering
\begin{tabular}{C{2cm}C{2cm}C{2cm}} 
\toprule
$B_f'/B_f$ & $\Lambda_{\DM}$ & $\rho$    \\ 
\hline
1.29 & 0.77 GeV & $\lesssim 100$ \\
\bottomrule
\end{tabular}
\caption{The results for the neutron portal. $B_f'/B_f$ is the final ratio of $B'$ to $B$ following the asymmetry transfer, $\Lambda_{\DM}$ is the dark confinement scale then required to reproduce the 5:1 ratio of dark and visible mass densities, and $\rho$ gives the possible ratios of electroweak scales for which this value of $\Lambda_{\DM}$ is achievable.}
\label{table:results_np}
\end{table}

\subsection{The lepton portal}
The effective interaction mediating this portal involves a lepton doublet and a Higgs doublet from each sector, and is given by
\begin{equation}\label{eq:lepport}
    \frac{1}{M_{ab}}\bar{l}_{iL}\Phi_a^cl_{jR}'\Phi_b' + H.c.
\end{equation}
where the indices $i, j = 1, 2, 3$ and $a, b = 1, 2$ specify the lepton generation and Higgs doublet number, respectively. 

These interactions allow for neutrino mass terms after electroweak symmetry breaking; the Higgs doublets gain VEVs given by $\ev{\Phi_a} = v_a/\sqrt{2}$ and $\ev{\Phi_b'} = w_b/\sqrt{2}$ and a mass term is produced with
\begin{equation}\label{eq:Mmnu}
    m_\nu = \frac{v_aw_b}{2M_{ab}}.
\end{equation}
The observational data for neutrino masses give an upper bound of $m_{\nu} \lesssim 0.12$ eV \cite{Aghanim:2018eyx}. This translates to a lower bound on $M_{ab}$ given by
\begin{equation}\label{eq:Mlimit}
    M_{ab} \gtrsim \frac{v_aw_b}{0.5 \mathrm{eV}}
\end{equation}

\subsubsection{Thermal equilibrium}

The approximate rate of the lepton portal interaction is $\Gamma \sim T^3/M_{ab}^2$. $\Gamma > H$ then implies that a given lepton portal is in thermal equilibrium for $T > {M_{ab}}^2/m_{\mathrm{Pl}}$. Combining this with Eqs.~\ref{eq:Mmnu} and \ref{eq:Mlimit}, we recast the condition as 
\begin{equation}
    T \gtrsim 0.25\left(\frac{v_aw_b}{\mathrm{GeV}^2}\right)^2\mathrm{GeV}.
\end{equation}

Thus, the temperature range for which a given lepton portal is in thermal equilibrium depends on which Higgs doublets take part in the interaction. Consider the lepton portal involving $\Phi_1$ and $\Phi_2'$ -- the two doublets which gain large VEVs $v_1 \approx v$ and $w_2 \approx w$. For a given value of $\rho$, we have $v = 246$ GeV and $w = 246\rho$ GeV, so this lepton portal is only in thermal equilibrium for $T > 10^9\rho^2$ GeV. This is a temperature range well above the dEWPT temperature, and so this lepton portal operator cannot serve to reprocess the asymmetries.

The other Higgs doublets in each sector have VEVs that are much smaller than $v$ and $w$, typically on the order of at most 1 GeV. Thus, ignoring Boltzmann suppression, the lepton portal involving $\Phi_2$ and $\Phi_1'$ remains in thermal equilibrium down to at least 1 GeV -- well below the temperature ranges we consider. However, these doublets are comprised of the heavy additional scalars, which become Boltzmann suppressed at high temperatures near the dark electroweak phase transition. For this portal to remain in thermal equilibrium long enough to reprocess the particle asymmetries between the sectors, we must work at a point in parameter space where the additional scalars have masses lower than the dEWPT temperature. In Sec.~\ref{sec:EWPT}, we found such parameter selections; for example, see the parameter point for region 3 in Table~\ref{table:samplepoints}. We analyse the asymmetry transfer at this parameter point, and thus work in an approximate temperature range between the critical temperature, $T_C = 1.7 \times 10^4$ GeV, and the mass of the heaviest additional scalar, $M_H = 8.8 \times 10^3$ GeV.

\subsubsection{Equilibrium conditions}

The lepton portal in Eq.~\ref{eq:lepport} induces chemical potential constraints given by
\begin{equation}\label{eq:lepton-portal-constraints}
    \lL + \mphiv - \mu_{\nu_{jR}'} - \mpOd = 0, \quad \lL + \mphiv - \mu_{e_{jR}'} - \mppd = 0.
\end{equation}

As we are also working in a temperature regime below $10^5$ GeV where all Yukawa interactions are in equilibrium, the discussion of the additional constraints in the visible and dark sectors is very similar to the neutron portal. The only difference is that the additional scalars are in equilibrium while the portals are active, and so we must set $N_{\Phi} = N_{\Phi'} = 2$ in the visible hypercharge and dark charge neutrality conditions. As we must have at least one $\mathcal{O}(1)$ dark Yukawa coupling, we consider a case where the dark top quark is Boltzmann suppressed in our temperature region of interest. We alter the charge neutrality and QCD sphaleron conditions as in the neutron portal case.

Recall that the visible lepton mass eigenstates are not mirror partners of the dark lepton mass eigenstates. So, a lepton portal respecting the mirror symmetry will induce interactions between all pairs of visible and dark lepton mass eigenstates. As we assign chemical potentials to the mass eigenstates in each sector, the lepton portal thus introduces all possible constraints of the form given in Eq.~\ref{eq:lepton-portal-constraints}. This has the effect of setting the chemical potentials to be equal for all left-handed visible leptons and for all right-handed dark leptons; that is, $\mu_{l_{1L}} = \mu_{l_{2L}} = \mu_{l_{3L}} \equiv \mu_{l_L}$, $\mu_{e_{1R}'} = \mu_{e_{2R}'} = \mu_{e_{3R}'} \equiv \mu_{e_{R}'}$ and $\mu_{\nu_{1R}'} = \mu_{\nu_{2R}'} = \mu_{\nu_{3R}'} \equiv \mu_{\nu_{R}'}$. This then sets $L_1$ = $L_2$ = $L_3$ = $L/3$ and $L_1'$ = $L_2'$ = $L_3'$ = $L'/3$; thus, when the lepton portal is active, there are only two conserved charges:
\begin{equation}
    \mathcal{L}_1 = B - L - L', \quad \mathcal{L}_2 = B'.
\end{equation}

As before, we define the visible and dark particle numbers as linear combinations of the conserved charges,
\begin{equation}
B = \sum_{i=1}^2a_i\mathcal{L}_i,
\quad
L = \sum_{i=1}^2b_i\mathcal{L}_i,
\quad
B' = \sum_{i=1}^2c_i\mathcal{L}_i,
\quad
L' = \sum_{i=1}^2d_i\mathcal{L}_i.
\end{equation}

We define $X$ and $Y$ as the initial asymmetries in $B'$ and $L'$ respectively, leading to the conserved charges $\mathcal{L}_1 = -Y$ and $\mathcal{L}_2 = X$. The same behaviour occurs as before with the final visible baryon asymmetry depending on the $B - L$ asymmetry transferred to the visible sector, and so we obtain
\begin{equation}
    \frac{B_f'}{B_f} = \frac{79}{28}\frac{-c_1Y + c_2X}{-a_1Y + a_2X + b_1Y - b_2X}.
\end{equation}

\subsubsection{Results}

For this section we give results as the range of relative sizes between $X$ and $Y$ that produce appropriate baryon ratios. To roughly identify the allowed range of baryon ratios, we again look to Fig.~1 from the original paper \cite{Lonsdale:2018xwd}. We are working at a parameter point where $\rho = 300$, and the heaviest of the new Yukawa couplings is on the order of the standard model top quark Yukawa coupling. Thus, we have a value of $\Lambda_{\DM}$ between $0.9$ and $2$ GeV. To reproduce the 5:1 ratio between the dark and visible mass densities, from Eq.~\ref{Eq:5ratio} it follows that we need a value for the baryon ratio satisfying $0.5 \lesssim B_f'/B_f \lesssim 1.1$. 

The asymmetry transfer by our lepton portal produces a valid baryon ratio for
\begin{equation}
    -3.6 < \frac{Y}{X} < -1.6,
\end{equation}
where $Y/X$ is the ratio of the initial dark lepton and baryon asymmetries. Thus we favour a dark EWBG which generates a lepton asymmetry of slightly larger magnitude than and opposite sign to the baryon asymmetry. While we have not provided a detailed calculation of the initial asymmetry generation dynamics in this work, to achieve similar initial asymmetries we expect that the heaviest dark quarks and leptons should have masses of similar sizes. This then requires at least one $\mathcal{O}(1)$ dark lepton Yukawa coupling. To make any more precise statements about the viability of this portal would require a detailed analysis of the dark electroweak baryogenesis dynamics.

\section{Dark Radiation and the Neutron Portal}\label{sec:darkrad}
So far, we have investigated the transfer of particle number asymmetries at temperatures well above the visible electroweak scale, providing a general analysis of effective operators which preserve a total particle number. Of these, the neutron portal was especially promising, as it could also naturally allow for the stringent bounds on dark radiation to be alleviated. However, to serve this dual role the neutron portal must remain in equilibrium until a decoupling temperature $T_\mathrm{dec}$ that lies between the visible and dark quark-hadron phase transition (QHPT) temperatures; in addition, this must be achieved in a way that does not introduce excessive fine-tuning to the extent that the model can no longer serve as a natural explanation of the cosmological coincidence $\Omega_{\DM} \simeq 5 \Omega_{\VM}$. In this section we discuss the difficulties that arise when attempting to implement the neutron portal at these low temperatures, in particular: (i) specifying a valid UV completion of the neutron portal and (ii) tracking the asymmetry transfer over a larger temperature range.

\subsection{UV-completing the neutron portal}
For the neutron portal to be active below the dark QHPT, it must remain in thermal equilibrium down to a temperature below $\Lambda_{\mathrm{DM}}$. While the specific value of $\Lambda_{\mathrm{DM}}$ depends on the spectrum of dark quark masses, it is at most a few GeV. From Eq.~\ref{Eq:neutron-portal-temp-range}, the neutron portal effective operator falls out of equilibrium at $T \approx (M^{10}/m_\mathrm{Pl})^{1/9}$. So, for the neutron portal to remain in operation at $T \sim 1$ GeV, we require a cutoff scale $M \lesssim 63$ GeV. At temperatures above this scale the effective operator description will be invalid, and so to properly analyse the asymmetry transfer we must provide a renormalisable realisation of the neutron portal effective operator.

A UV completion for this operator was given in Ref. \cite{Berezhiani:2005hv}, and a similar interaction was given in Ref. \cite{Gu:2011ff} in the context of neutron-antineutron oscillation. Similarly to these papers, we introduce a scalar diquark $S \sim (\bm{3},\bm{1},\frac{2}{3})$ with baryon number $B = -\frac{2}{3}$ and a gauge singlet fermion $N_R \sim (\bm{1},\bm{1},0)$ with baryon number $B = -1$, as well as their mirror partners $S'$ and $N_L'$. Assuming $B - B'$ conservation, the new Yukawa and mass terms are given by
\begin{equation}
\begin{split}
    \Lagr &\supset \lambda_i(S\bar{u}_{Ri}N_R^c + S'\bar{u}_{Li}'N_L^{c\prime}) + \kappa_{ij}(S\bar{d^c}_{Ri}d_{Rj} + S'\bar{d^c}_{Li}'d_{Lj}') \\
    &+ M_S^2(S^{\ast}S + S^{\ast\prime}S') + M_NN_RN_L'.
\end{split}
\end{equation}

$B - B'$ conservation forbids Majorana mass terms for the singlet fermions, preventing the washout of any dark or visible baryon number asymmetry carried by the respective singlet fermion. While the mirror symmetry gives equal mass terms for $S$ and $S'$, they can obtain differing masses following symmetry breaking through their couplings to the Higgs doublets. In the ASB limit, the relevant couplings are given by
\begin{equation}
\begin{split}
    \Lagr &\supset \eta_1(S^{\ast}S\Phi_1^\dagger\Phi_1 + S'^{\ast}S'\Phi_1'^\dagger\Phi_1') + \eta_2(S^{\ast}S\Phi_2^\dagger\Phi_2 + S'^{\ast}S'\Phi_2'^\dagger\Phi_2')\\ 
    &+ \eta_3(S^{\ast}S\Phi_1'^\dagger\Phi_1' + S'^{\ast}S'\Phi_1^\dagger\Phi_1) + \eta_4(S^{\ast}S\Phi_2'^\dagger\Phi_2' + S'^{\ast}S'\Phi_2^\dagger\Phi_2),
\end{split}
\end{equation}
producing scalar diquark masses
\begin{equation}
\begin{split}
    m_S^2 = M_S^2 + \frac{v^2}{2}(\eta_1 + \rho^2\eta_4),\\
    m_{S'}^2 = M_S^2 + \frac{v^2}{2}(\eta_3 + \rho^2\eta_2).
\end{split}
\end{equation}

\begin{figure}
\centering
\includegraphics[]{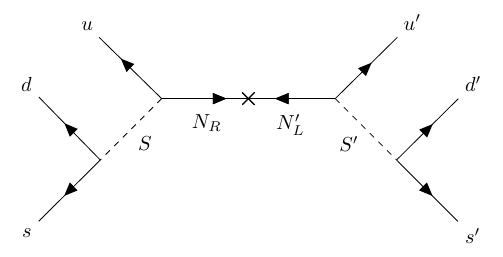}
\caption{\label{fig:neutron-portal-interaction}Diagram inducing the neutron portal effective operator for temperatures $T < m_S, m_{S'}, M_N$.}
\end{figure}

The neutron portal operators are induced by diagrams such as that given in Fig.~\ref{fig:neutron-portal-interaction}. As before, to ensure the stability of the dark neutron -- our dark matter candidate -- we cannot allow the neutron portal to involve only the lightest dark quarks, $u'$ and $d'$. We thus need to introduce some additional flavour structure to the Yukawa interactions, such that one of $\lambda_1$ or $\kappa_{11}$ is equal to zero. 

At temperatures $T < m_S, m_{S'}, M_N$, and assuming $\bigO(1)$ Yukawa couplings, the cutoff scale for the neutron portal effective operator is given by $M \sim (m_S^2m_{S'}^2M_N)^{\frac{1}{5}}$. Visible scalar diquarks at masses below a few TeV have been disfavoured by collider searches \cite{Sirunyan:2018xlo}, so we take $m_S \sim 10^4$ GeV, which can be easily achieved with $\rho \sim 100$ and $\bigO(1)$ values of $\eta_4$. Then, for the cutoff scale to satisfy $M \lesssim 63$ GeV, we require $m_{S'}^2M_N \lesssim 10$ GeV$^3$. For $m_{S'}^2 \gtrsim 10$ GeV$^2$ we then require $M_N \lesssim 1$ GeV, which means that at the dark QHPT $N_R$ and $N_L'$ remain in equilibrium and the neutron portal is not described by a single portal operator. Achieving $m_{S'}^2 \lesssim 10$ GeV$^2$ is not feasible without significant fine-tuning, requiring $\eta_3 \lesssim 10^{-3}$ and $\eta_2 \lesssim 10^{-7}$ for $\rho \sim 100$. Thus, we do have to consider the situation where $N_R$ and $N_L'$ have masses smaller than the dark QHPT temperature $T \sim 1$ GeV.

\begin{figure}
\includegraphics[]{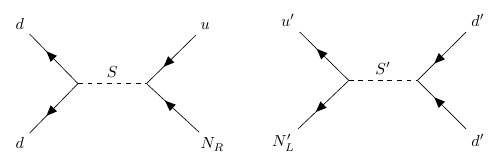}
\caption{\label{fig:S-mediated-interactions}Diagrams inducing effective operators for temperatures given by $M_N < T < m_S, m_{S'}$.}
\end{figure}

In this case, for the neutron portal to be active at the dark QHPT, we need the effective operators induced by the $S$-/$S'$-mediated interactions of Fig.~\ref{fig:S-mediated-interactions} to be in thermal equilibrium at $T \sim1$ GeV. Assuming $\bigO(1)$ Yukawa couplings, these operators are in thermal equilibrium for $T \gtrsim (m_{S^{(\prime)}}^4/m_{\mathrm{Pl}})^{1/3} = 200$ MeV assuming $m_S, m_{S'} \sim 10^4$ GeV. So, the given completion for the neutron portal allows it to be active at the dark QHPT temperature for $m_S, m_{S'} \sim 10^4$ GeV and $M_N$ small enough for $N_R$ and $N_L'$ to remain in thermal equilibrium.

However, this situation introduces an issue for the stability of our dark matter candidate, the dark neutron: if $M_N$ is lower than $m_{n'}$, then the decay mode $n' \rightarrow N_L'\gamma'$ becomes available. The dark neutron mass is a factor of a few times $\Lambda_{\mathrm{QCD}}$, which also approximately gives the dark QHPT temperature. So, if $M_N$ is smaller than the dark QHPT temperature to allow the neutron portal to decouple between the visible and dark QHPTs, then it will be lighter than $n'$ and the dark matter will not be stable. The flavour structure of the Yukawa couplings does prevent this decay occurring at tree-level; however, the kinematically-allowed decay can still occur at one-loop level.

This instability could be avoided if $M_N$ is greater than $m_n'$, at a value around a few GeV, and if the singlet fermions do not fall out of thermal equilibrium until a temperature a factor of 10 or so smaller than their mass. While possible, this places quite a tight restriction on the $M_N$, as it must be only slightly higher than $m_n'$ to be able to remain in thermal equilibrium down to a temperature between the visible and dark QHPT temperatures. If this is the case, however, then the neutron portal can remain active down to the desired decoupling temperature.

\subsection{Asymmetry transfer}
We now analyse the asymmetry transfer due to this specific neutron portal. As this cross-sector interaction is active from the dEWPT at around $10^5$ GeV down to the decoupling temperature between the visible and dark QHPTS at around $1$ GeV, determining the final baryon ratio $B'/B$ is more complicated than the cases considered in Sec.~\ref{sec:asym}. 

Recall that at a given temperature, our process is to (i) identify which particle species are in equilibrium and assign them chemical potentials, (ii) identify the interactions in thermal equilibrium that constrain these chemical potentials, (iii) identify the conserved charges that correspond to the remaining free parameters, and (iv) solve for the chemical potentials in terms of the initial conditions on these conserved charges. Since the neutron portal is now active over a large temperature range, the chemical potentials reequilibrate and new charges become conserved as various particle species and interactions fall out of equilibrium. 

To account for this reequilibration, we first identify the new conserved charges as well as the temperatures at which they begin to be conserved following the freeze-out of particular particle species and interactions. When each new charge begins to be conserved, we can calculate its asymmetry in terms of the other conserved charges at that temperature. So, beginning at the dEWPT temperature $T \sim 10^5$ GeV with initial asymmetries given by $B' = X$ and $L' = Y$, we can calculate each new conserved charge in terms of $X$ and $Y$. Continuing this process down to $1$ GeV, we obtain the final baryon ratio $B'/B$ immediately prior to the dark QHPT.

Although the neutron portal freezes out between the visible and dark QHPT temperatures due to the Boltzmann suppression of the quarks involved, we do not continue to track the baryon ratio after the dark QHPT commences. This is due to the nonperturbative strong dynamics of the dark QHPT, which cannot be handled by our approximate calculation method. While this introduces additional uncertainty to our calculations, our goal is not to calculate a precise ratio, but just to show that a reasonable ratio of $B'/B < 5$ can be obtained by the neutron portal operator acting until low temperatures. Additionally, given that the neutron portal freezes out shortly after the dark QHPT commences, we claim that the final baryon number ratio should not change after the transition by more than a factor of a few.

\subsubsection{Conserved charges}

To simplify the analysis further, we will drop generational indices from our chemical potentials, setting equal chemical potentials for the particles of each type in thermal equilibrium. We first work at a temperature on the order of $10^4$ GeV where the scalar diquarks $S$ and $S'$ have frozen out but the dark Higgs boson $h'$ is still in equilibrium. The particle species and interactions in equilibrium are then the same as in Sec.~\ref{sec:neutron-portal}, but with the neutron portal constraint replaced by constraints from the $S$- and $S'$-mediated effective operators
\begin{equation}
    \mu_{u_R} + \mu_{N_R} + 2\mu_{d_R} = 0, \quad \mu_{u_L'} + \mu_{N_L'} + 2\mu_{d_L'} = 0
\end{equation}
and the gauge singlet mass term setting $\mu_{N_R} = \mu_{N_l'}$. Then there are only two conserved charges,
\begin{equation}
    \mathcal{L}_1 = B + B' - L, \quad \mathcal{L}_2 = L', 
\end{equation}
with initial conditions $\mathcal{L}_1 = X$ and $\mathcal{L}_2 = Y$. 

After the dark Higgs boson $h'$ freezes out at around $10^4$ GeV, the dark Yukawa interactions are replaced by four-fermion interactions mediated by the dark Higgs doublets,
\begin{equation}\label{Eq:higgs-mediated-yukawas-dark}
    \mu_{u_R'} - \mu_{u_L'}  + \mu_{e_R'} - \mu_{e_L'} = 0, \quad \mu_{u_R'} - \mu_{u_L'}  + \mu_{d_R'} - \mu_{d_L'} = 0, \quad \mu_{d_R'} - \mu_{d_L'}  - \mu_{e_R'} + \mu_{e_L'} = 0,
\end{equation}
and there is a new conserved charge given by $\mathcal{L}_3 = \mu_{u_R'} - \mu_{d_R'} + \mu_{\nu_R'} - \mu_{e_R'}$; that is, $u_R'$ and $\nu_R'$ have $\mathcal{L}_3$ charge 1 and $d_R'$ and $e_R'$ have $\mathcal{L}_3$ charge -1.

These Higgs-mediated interactions remain in equilibrium for temperatures given by $T > \left((m_{h'}w)^4/(4{m_1}^2{m_2}^2m_{\mathrm{Pl}})\right)^{1/3}$, where $m_1$ and $m_2$ are the masses of the fermion species involved and we have used the Cheng-Sher \emph{Ansatz} \cite{Cheng:1987rs} for the dark Yukawa couplings. 

We work at a benchmark scenario with $\rho \simeq 100$, $z_2 \simeq 0.025$, and selected dark quark masses of $m_{c'} \simeq 50$ GeV, $m_{\mu'} \simeq 50$ GeV, and $m_{b'} \simeq 500$ GeV. Then the Higgs-mediated interaction constraints of Eq.~\ref{Eq:higgs-mediated-yukawas-dark} apply until $T \simeq 60$ GeV (for $c_L' + c_R' \leftrightarrow \mu_L' + \mu_R'$), $T \simeq 500$ GeV (for $c_L' + c_R' \leftrightarrow b_L' + b_R'$), and $T \simeq 500$ GeV (for $b_L' + b_R' \leftrightarrow \mu_L' + \mu_R'$), respectively.

So, after the constraints $\mu_{u_R'} - \mu_{u_L'}  + \mu_{d_R'} - \mu_{d_L'} = 0$ and $\mu_{d_R'} - \mu_{d_L'}  - \mu_{e_R'} + \mu_{e_L'} = 0$ freeze out at $T \sim 500$ GeV, the conserved charge $\mathcal{L}_3 = \mu_{u_R'} - \mu_{d_R'} + \mu_{\nu_R'} - \mu_{e_R'}$ splits into two conserved charges $\mathcal{L}_3 = \mu_{u_R'} - \mu_{e_R'}$ and $\mathcal{L}_4 = \mu_{d_R'} - \mu_{\nu_R'}$.

The next stage is shortly after the vEWPT at the electroweak sphaleron freeze out temperature $T \sim 150$ GeV. This assumes the vEWPT is crossover as it is in the SM \cite{Gurtler:1997kh}; we make this assumption since the dynamics of the vEWPT are controlled by the couplings of $\Phi_1$ which are very similar to that of the SM Higgs doublet. After this point, the charge $\mathcal{L}_1 = B + B' - L$ splits into two conserved charges, $\mathcal{L}_1 = B + B'$ and $\mathcal{L}_5 = L$. 

After the freeze out of the visible Higgs around its mass of $125$ GeV, there is a new conserved charge $\mathcal{L}_6 = \mu_{u_L} - \mu_{d_L} + \mu_{\nu_L} - \mu_{e_L}$. The final new conserved charge is $\mathcal{L}_7 = \mu_{u_R'} + \mu_{d_R'}$, which becomes conserved following the freeze out of the dark Higgs-mediated constraint $\mu_{u_R'} - \mu_{u_L'}  + \mu_{e_R'} - \mu_{e_L'} = 0$ at $60$ GeV. A summary of these conserved charges and the temperatures at which they are first conserved is presented in Table~\ref{table:conserved-charges}.

\begin{table}
\centering
\begin{tabular}{C{2cm}l}
\toprule
$T$ [GeV] & \textbf{Conserved Charges}     \\ 
\hline
$10^5$ & $\{B + B' - L'$, $L'\}$ \\
$10^4$ & $\{B + B' - L'$, $L'$, $\mu_{u_R'} - \mu_{d_R'} + \mu_{\nu_R'} - \mu_{e_R'}\}$ \\
$500$ & $\{B + B' - L'$, $L'$, $\mu_{u_R'} - \mu_{e_R'}$, $\mu_{d_R'} - \mu_{\nu_R'}\}$ \\
$150$ & $\{B + B'$, $L'$, $\mu_{u_R'} - \mu_{e_R'}$, $\mu_{d_R'} - \mu_{\nu_R'}$, $L\}$ \\
$125$ & $\{B + B'$, $L'$, $\mu_{u_R'} - \mu_{e_R'}$, $\mu_{d_R'} - \mu_{\nu_R'}$, $L$, $\mu_{u_L} - \mu_{d_L} + \mu_{\nu_L} - \mu_{e_L}\}$ \\
$60$ & $\{B + B'$, $L'$, $\mu_{u_R'} - \mu_{e_R'}$, $\mu_{d_R'} - \mu_{\nu_R'}$, $L$, $\mu_{u_L} - \mu_{d_L} + \mu_{\nu_L} - \mu_{e_L}$, $\mu_{u_R'} + \mu_{d_R'}\}$ \\
\bottomrule
\end{tabular}
\caption{The set of conserved charges $\{\mathcal{L}_i\}$ along with the approximate temperature $T$ at which each new charge first becomes conserved.}\label{table:conserved-charges}
\end{table}

\subsubsection{Results}
At around $1$ GeV, just before the dark QHPT commences, we can calculate $B'$ and $B$ in terms of the seven conserved charges. Starting with the initial conditions $\mathcal{L}_1 = X$ and $\mathcal{L}_2$, we determine each new conserved charge in terms of $X$ and $Y$; continuing all the way to $1$ GeV, we obtain
\begin{equation}
\begin{split}
    B'_f &\simeq 0.28X + 0.07Y\\
    B_f &\simeq 0.24X - 0.003Y.
\end{split}
\end{equation}

Consider a case where no dark lepton asymmetry is generate during dark EWBG and thus $Y = 0$; this could easily arise if the Yukawa couplings of the dark leptons are all smaller than $\mathcal{O}(1)$, as they are in the SM. Then, the final ratio of baryon numbers is given by
\begin{equation}
    \frac{B'_f}{B_f}|_{Y = 0} \simeq 1.1
\end{equation}

So, the neutron portal scenario we consider can naturally generate similar asymmetries in visible and dark baryon number. We also check when the constraint $B'/B < 5$ applies in terms of the relative sizes of the initial dark baryon and lepton asymmetries, finding that is it satisfied for $Y/X \lesssim 11.4$. Thus, we can remain fairly agnostic about the specifics of the asymmetry generation through dark EWBG, as a reasonable and related baryon ratio can be obtained for an initial lepton asymmetry up to an order of magnitude larger than the initial baryon asymmetry.

\section{Conclusion}\label{sec:conclusion}
The 5:1 ratio between the present-day mass densities of dark matter and visible matter is one of the few tantalising hints we have toward the fundamental nature of dark matter. This apparent coincidence of both cosmological number densities and mass scales between dark and visible relic species suggests a deep link between the two forms of matter, motivating the search for a comprehensive dark matter model where this relationship arises naturally.

While asymmetric dark matter models provide a variety of ways to relate the number densities of visible and dark matter, relating the particle masses presents a challenge that is more difficult and thus less frequently addressed. In this work we focused on extending the mirror two Higgs doublet model of Ref.~\cite{Lonsdale:2018xwd}, where the dark matter consists of neutrons of a dark QCD whose confinement scale is related to $\Lambda_{\mathrm{QCD}}$ by a discrete symmetry that is spontaneously broken at a high scale. While this earlier work generated related visible and dark baryon number densities through thermal leptogenesis, we sought to implement electroweak baryogenesis at the dark electroweak phase transition as the method for generating a particle asymmetry. 

In this work we did not present a fully detailed theory of electroweak baryogenesis; rather, we completed some preliminary steps to demonstrate the feasibility of such a model, and to show that it could be naturally realised within the mirror 2HDM framework of Ref.~\cite{Lonsdale:2018xwd}. We first showed in Sec.~\ref{sec:EWPT} that for a number of regions of parameter space the dark electroweak phase transition is strongly first-order, as is necessary to provide the out-of-equilibrium dynamics in EWBG. In Sec.~\ref{sec:asym} we then considered the generation and reprocessing of the dark baryon asymmetry generated through dark EWBG by cross-sector effective operator interactions. For both interactions we analysed -- the neutron portal and the lepton portal -- the final visible and dark baryon number densities obtained were of a similar order. However, in the case of the lepton portal, the present-day baryon asymmetry ratio depended on the relative sizes of the dark lepton and baryon asymmetries generated at the dEWPT. Determining these initial conditions requires a full calculation of the EWBG dynamics. 

In addition to providing the initial conditions for the lepton portal, a full EWBG calculation would also show whether a sufficiently large dark baryon asymmetry can be generated; that is, large enough that the correct visible baryon number density is reproduced following the asymmetry transfer. Using a rough condition we showed that, given at least one $\mathcal{O}(1)$ dark quark Yukawa coupling, a sufficiently large baryon asymmetry should be able to be generated through $CP$ violation in the dark CKM matrix. A quantitative EWBG analysis is necessary to turn this work into a complete theory.

Lastly in Sec.~\ref{sec:darkrad} we considered the tricky issue of alleviating the BBN bounds on dark radiation, which presents a common challenge in ADM theories. The most promising and natural solution given in the original paper was a neutron portal operator holding the visible and dark sectors in thermal equilibrium until a point shortly after the dark quark-hadron phase transition. The notion of the neutron portal serving a dual role by transferring asymmetries and addressing the dark radiation issue is greatly appealing; however, we showed it is difficult to implement the neutron portal down to temperatures around $1$ GeV. In particular, we identified a UV completion that allowed the neutron portal to operate successfully, but only if the gauge singlet fermions $N_R$ and $N_L'$ have a mass just larger than the dark neutron. This is quite a tight restriction, and presents an unwanted source of fine-tuning. Given this UV completion, we then calculated the asymmetry transfer -- noting the large uncertainty introduced by the non-perturbative dynamics of the dark QHPT -- and showed that the neutron portal can generate visible and dark baryon asymmetries of a similar order while also helping to obey the dark radiation constraints.

\section*{Acknowledgements}
We thank Stephen Lonsdale for helpful correspondence. This work was supported in part by the Australian Research Council and the Australian Government Research Training Program Scholarship initiative.

\end{document}